\documentclass[preprint,preprintnumbers,showpacs,aps,prd,amssymb]{revtex4-1}

\usepackage{graphicx}
\usepackage{bm}
\usepackage{amsmath}
\usepackage{xcolor}

\def\calH{{\cal H}}

\def\calN{{\cal N}}
\def\calO{{\cal O}}

\def\hbar{{\bar h}}

\def\sbar{{\bar s}}

\def\nubar{{\bar\nu}}

\def\SM{{\rm SM}}
\def\exp{{\rm exp}}
\def\NP{{\rm NP}}
\def\GeV{{\rm GeV}}
\def\TeV{{\rm TeV}}
\def\Br{{\rm Br}}
\def\eff{{\rm eff}}
\def\min{{\rm min}}
\def\dof{{\rm d.o.f.}}
\def\best{{\rm best}}

\def\nn{\nonumber}

\def\Ks{{K^{(*)}}}
\def\mhat{{\hat m}}
\def\shat{{\hat s}}
\def\uhat{{\hat u}}

\def\B2Ksll{{B\to\Ks\ell^+\ell^-}}
\def\Bs2mumu{{B_s\to\mu^+\mu^-}}
\def\B2Knunu{{B^+\to K^+\nu\nubar}}
\def\P5p{{P_5'(B^+\to K^{*+}\mu^+\mu^-)}}

\def\perfb{{\rm fb}^{-1}}

\begin{document}
\title{New Physics effects in $R(\Ks)$, $\Bs2mumu$, and $\B2Knunu$}
\author{Jong-Phil Lee}
\email{jongphil7@gmail.com}
\affiliation{Sang-Huh College,
Konkuk University, Seoul 05029, Korea}

\begin{abstract}
We analyze possible new physics (NP) effects on $b\to s$ transition processes, 
$R(\Ks)$, $\Bs2mumu$, and $\B2Knunu$ decays.
Though recent data for $R(\Ks)$ and $\Br(\Bs2mumu)$ are compatible with the standard model (SM),
there are still rooms for NP beyond the SM.
Especially $\Br(\B2Knunu)$ is measured to exceed the theoretical predictions.
We parameterize the NP effects in a generic way with explicit NP scale $M_\NP$ and
some possible powers of it.
For reasonable ranges of NP fermionic couplings we find a window of 
$3.17~\TeV\le M_\NP\le14.9~\TeV$ for new particles.
When including $\Br(B^+\to K^+\mu^+\mu^-)$ and the angular observable $\P5p$
we have a wider range of the window.
Implications of our analysis about specific NP models are discussed.
\end{abstract}

\maketitle
\section{Introduction}
%
%
%
Flavor changing neutral currents (FCNCs) are very important to the standard model (SM) of particle physics
because they provide a strong and useful testbed for new physics (NP) beyond the SM.
In the SM, FCNC processes are induced by loop and further suppressed by CKM factors and thus
very sensitive to NP contributions.
\par
Until recently, the ratio of the branching ratios
\begin{equation}
R(\Ks)\equiv\frac{\Br(B\to\Ks\mu^+\mu^-)}{\Br(B\to\Ks e^+e^-)}~,
\end{equation}
showed some deviations from the SM, casting a puzzle of lepton universality violation
\cite{LHCb1705,LHCb2103,Geng2103}.
But new experimental data from the LHCb \cite{LHCb2212_52,LHCb2212_53}
\begin{eqnarray}
\label{RKsLHCb}
R(K)_L &=& 0.994^{+0.094}_{-0.087}~,~~~
R(K)_C = 0.949^{+0.048}_{-0.047}~,\\
R(K^*)_L &=& 0.927^{+0.099}_{-0.093}~,~~~
R(K^*)_C = 1.027^{+0.077}_{-0.073}~,
\end{eqnarray}
where $L(C)$ means the low (central) $q^2$ region, $[0.1,1.1]~ \GeV^2$ ($[1.1,6.0] ~\GeV^2$),
turned out to be consistent with the SM predictions.
The SM calculations for the $R(\Ks)$ are very close to unity \cite{Hiller0310,Bobeth0709,Geng1704,Bordone1605}
\begin{eqnarray}
R(K)_\SM[1.0,6.0] &=&1.0004^{+0.0008}_{-0.0007}~, \nn\\
R(K^*)_\SM[0.1,1.1] &=& 0.983\pm 0.014~, \nn\\
R(K^*)_\SM[1.1,6.0] &=& 0.996^{+0.002}_{-0.002}~.
\label{RKs_SM}
\end{eqnarray}
Also the CMS collaboration measured $R(K)_C$ as \cite{CMS2023}
\begin{equation}
R(K)_C=0.78^{+0.47}_{-0.23}~,
\end{equation}
which is compatible with the SM.
Closely related process is $B_s\to\mu^+\mu^-$ decay.
The world average of the experimental branching ratio is \cite{PDG2024}
\begin{equation}
\Br(\Bs2mumu)_\exp = (3.34\pm0.27)\times 10^{-9}~,
\end{equation}
which is in agreement with the SM result of \cite{Czaja2407}
\begin{equation}
\Br(\Bs2mumu)_\SM = (3.64\pm0.12)\times 10^{-9}~.
\end{equation}
\par
On the other hand, Belle II Collaboration recently measured $B\to K\nu\nu$ branching ratios as \cite{Belle2_2311}
\begin{equation}
\Br(\B2Knunu)_\exp = (2.3\pm 0.7)\times 10^{-5}~,
\end{equation}
which exceeds the SM predictions \cite{Becirevic2301}
\begin{equation}
\Br(\B2Knunu)_\SM = (4.43\pm0.31)\times 10^{-6}~.
\end{equation}
\par
Thus the NP effects, if exist, are suppressed in $R(\Ks)$ and $\Br(\Bs2mumu)$ 
while enhanced in $\Br(\B2Knunu)$.
It would be quite interesting to deal with both opposite trends simultaneously.
There have been many studies of NP effects on $b\to s$ transition processes:
supersymmetry (SUSY) \cite{Altmannshofer2002,Bardhan2107,Zheng2410}, 
the leptoquark (LQ) \cite{Hiller1408,Dorsner1603,Bauer1511,Chen1703,Crivellin1703,Calibbi1709,Blanke1801,Nomura2104,Angelescu2103,Du2104,Cheung2204}, 
$Z'$ model \cite{Crivellin1501,Crivellin1503,Chiang1706,King1706,Chivukula1706,Cen2104,Davighi2105}, 
new scalar models \cite{Hu1612,Crivellin1903,Rose1903,Ho2401}, and
unparticles \cite{JPL2106}, etc.
\par
In previous works of \cite{JPL2110,JPL2208} we analyzed $R(\Ks)$ in a more generic way
where NP effects are encoded in the relevant Wilson coefficients as $C_j^\NP\sim(v/M_\NP)^\alpha$.
Here $v$ is the SM vacuum expectation value and $M_\NP$ is the NP scale.
Inspired by unparticle scenario, $\alpha$ can be a non-integer parameter.
Ordinary NP particles would contribute with $\alpha=2$ at tree level.
Since $R(\Ks)$ and $\Br(\Bs2mumu)$ tend to suppress NP effects, 
they would impose lower bounds on $M_\NP$ assuming reasonably finite fermionic couplings.
On the contrary $\Br(\B2Knunu)$ favors finite NP effects so that it would also constrain upper bounds of $M_\NP$.
In this way we can estimate allowed windows of $M_\NP$ up to involved fermionic couplings.
One could easily check the validity of some NP models with this generic framework.
\par
The paper is organized as follows.
In the next Section the decay rates for $B\to\Ks\ell^+\ell^-$, $\Bs2mumu$, and $\B2Knunu$ are provided.
NP effects are parametrized as mentioned above.
Section III discusses our results and their physical meanings.
In Sec.\ IV we conclude.
%
%
%
\section{Decay rates}
%
%
%
The $b\to s\ell^+\ell^-$ transition can be described by the following effective Hamiltonian
\begin{equation}
\calH_{\rm eff}(b\to s\ell\ell) = 
-\frac{4G_F}{\sqrt{2}}V_{tb}V_{ts}^*\sum_i \left[
C_i(\mu)\calO_i(\mu)+C'_i(\mu)\calO'_i(\mu)\right]~.
\end{equation} 
For $R(\Ks)$ the relevant operators are \cite{Geng2103,Alonso14,Geng17}
\begin{eqnarray}
\calO_9 &=& \frac{e^2}{16\pi^2}\left(\sbar\gamma^\mu P_L b\right)\left({\bar\ell}\gamma_\mu\ell\right)~,\nn\\
\calO_{10} &=& \frac{e^2}{16\pi^2}\left(\sbar\gamma^\mu P_L b\right)\left({\bar\ell}\gamma_\mu\gamma_5\ell\right)~.
\label{O9O10}
\end{eqnarray}
The primed operators are proportional to the right-handed quarks, 
$\calO'_{9,10}\sim (\sbar\gamma^\mu P_R b)$.
In this analysis we do not consider $\calO'_{9,10}$ operators for simplicity.
One can expand the matrix elements for $B\to\Ks$ as \cite{Ali99}
\begin{eqnarray}
\langle K(p)|\sbar\gamma_\mu b|B(p_B)\rangle&=&
f_+\left[(p_B+p)_\mu-\frac{m_B^2-m_K^2}{s}q_\mu\right]+\frac{m_B^2-m_K^2}{s}f_0 q_\mu~,\\
\langle K(p)|\sbar\sigma_{\mu\nu} q^\nu(1+\gamma_5)b|B(p_B)\rangle&=&
i\left[(p_B+p)_\mu s -q_\mu(m_B^2-m_K^2)\right]\frac{f_T}{m_B+m_K}~,\\
\langle K^*(p)|(V-A)_\mu|B(p_B)\rangle&=&
-i\epsilon_\mu^*(m_B+m_{K^*})A_1+i(p_B+p)_\mu(\epsilon^*\cdot p_B)\frac{A_2}{m_B+m_{K^*}}\nn\\
&&
+iq_\mu(\epsilon^*\cdot p_B)\frac{2m_{K^*}}{s}(A_3-A_0)
+\frac{\epsilon_{\mu\nu\rho\sigma}\epsilon^{*\nu}p_B^\rho b^\sigma}{m_B+m_{K^*}}2V~,
\end{eqnarray}
where $f_{+,0,T}(s), ~A_{0,1,2}(s), ~T_{1,2,3}(s), V(s)$, and $f_-=(f_0-f_+)(1-\mhat_K^2)/\shat$ are the form factors.
Here,
\begin{eqnarray}
 q&=&p_B-p~,~~~s=q^2=(p_B-p)^2~,\\
 \shat&=&\frac{s}{m_B^2}~,~~~\mhat_i=\frac{m_i}{m_B}~.
 \end{eqnarray}
The form factors are functions of $s$ and we choose the exponential form as \cite{Ali99}
\begin{equation}
F(\shat)=F(0)\exp\Big(c_1\shat+c_2\shat^2+c_3\shat^3\Big)~,
\label{expFF}
\end{equation}
where $c_i$ are the related coefficients.
\par
The differential decay rates for $B\to\Ks\ell^+\ell^-$ with respect to $s$ are given by \cite{Chang2010}
\begin{eqnarray}
\frac{d\Gamma_K}{d\shat}&=&
\frac{G_F^2\alpha^2m_B^5}{2^{10}\pi^5}|V_{tb}V_{ts}^*|^2\uhat_{K,\ell}\left\{
(|A'|^2+|C'|^2)\left(\lambda_K-\frac{\uhat_{K,\ell}^2}{3}\right)
+|C'|^24\mhat_\ell^2(2+2\mhat_K^2-\shat) \right .\nn\\
&&\left. +{\rm Re}(C'D'^*)8\mhat_\ell^2(1-\mhat_K^2)+|D'|^24\mhat_\ell^2\shat\right\}~,
\\
\frac{d\Gamma_{K^*}}{d\shat}&=&
\frac{G_F^2\alpha^2m_B^5}{2^{10}\pi^5}|V_{tb}V_{ts}^*|^2\uhat_{K^*,\ell}\left\{
\frac{|A|^2}{3}\shat\lambda_{K^*}\left(1+\frac{2\mhat_\ell^2}{\shat}\right)
+|E|^2\shat\frac{\uhat_{K^*,\ell}^2}{3}\right.\nn\\
&&
+\frac{|B|^2}{4\mhat_{K^*}^2}\left[\lambda_{K^*}-\frac{\uhat_{K^*,\ell}t^2}{3}+8\mhat^2_{K^*}(\shat+2\mhat_\ell^2)\right]
+\frac{|F|^2}{4\mhat_{K^*}^2}\left[\lambda_{K^*}-\frac{\uhat_{K^*,\ell}^2}{3}+8\mhat_{K^*}^2(\shat-4\mhat_\ell^2)\right]\nn\\
&&
+\frac{\lambda_{K^*}|C|^2}{4\mhat_{K^*}^2}\left(\lambda_{K^*}-\frac{\uhat_{K^*,\ell}^2}{3}\right)
+\frac{\lambda|_{K^*}|G|^2}{4\mhat_{K^*}^2}\left[\lambda_{K^*}-\frac{\uhat_{K^*,\ell}^2}{3}
	+4\mhat_\ell^2(2+2\mhat_{K^*}^2-\shat)\right]\nn\\
&&
-\frac{{\rm Re}(BC^*)}{2\mhat_{K^*}^2}\left(\lambda_{K^*}-\frac{\uhat_{K^*,\ell}^2}{3}\right)(1-\mhat_{K^*}^2-\shat)\nn\\
&&
-\frac{{\rm Re}(FG^*)}{2\mhat_{K^*}^2}\left[\left(\lambda_{K^*}-\frac{\uhat_{K^*,\ell}^2}{3}\right)(1-\mhat_{K^*}^2
	-\shat)-4\mhat_\ell^2\lambda_{K^*}\right]\nn\\
&&\left.
-\frac{2\mhat_\ell^2}{\mhat_{K^*}^2}\lambda_{K^*}\left[{\rm Re}(FH^*)-{\rm Re}(GH^*)(1-\mhat_{K^*}^2)\right]
+\frac{\mhat_\ell^2}{\mhat_{K^*}^2}\shat\lambda_{K^*}|H|^2\right\}~,
\end{eqnarray}
where the kinematic variables are
\begin{eqnarray}
\lambda_H&=&1+\mhat_H^4+\shat^2-2\shat-2\mhat_H^2(1+\shat)~,\\
\uhat_{H,\ell}&=&\sqrt{\lambda_H\left(1-\frac{4\mhat_\ell^2}{\shat}\right)}~.
\end{eqnarray}
Here $A',\cdots, D'$ and $A,\cdots, H$ are the auxiliary functions.
They are defined by the form factors combined with the Wilson coefficients as \cite{Ali99},
%
\begin{eqnarray}
A'&=&C_9 f_+ +\frac{2\mhat_b}{1+\mhat_K}C_7^\eff f_T ~,\\
B'&=&C_9 f_- -\frac{2\mhat_b}{\shat}(1-\mhat_K)C_7^\eff f_T ~,\\
C'&=&C_{10} f_+ ~,\\
D'&=&C_{10} f_- ~,
\end{eqnarray}
and
\begin{eqnarray}
A&=&\frac{2}{1+\mhat_{K^*}}C_9 V+\frac{4\mhat_b}{\shat}C_7^\eff T_1~,\\
B&=&(1+\mhat_{K^*})\left[C_9 A_1+\frac{2\mhat_b}{\shat}(1-\mhat_{K^*})C_7^\eff T_2\right]~,\\ 
C&=&\frac{1}{1-\mhat_{K^*}^2}\left[(1-\mhat_{K^*})C_9 A_2
	+2\mhat_b C_7^\eff \left(T_3+\frac{1-\mhat_{K^*}^2}{\shat}T_2\right)\right]~,\\
D&=&\frac{1}{\shat}\left\{C_9\left[(1+\mhat_{K^*})A_1-(1-\mhat_{K^*})A_2
	-2\mhat_{K^*}A_0\right]-2\mhat_b C_7^\eff T_3\right\}~,\\
E&=&\frac{2}{1+\mhat_{K^*}}C_{10} V ~,\\
F&=&(1+\mhat_{K^*})C_{10} A_1 ~,\\
G&=&\frac{1}{1+\mhat_{K^*}}C_{10} A_2 ~,\\
H&=&\frac{1}{\shat}C_{10}\left[(1+\mhat_{K^*})A_1-(1-\mhat_{K^*})A_2-2\mhat_{K^*} A_0\right] ~.
\end{eqnarray}
%
\par
Possible NP effects on $R(\Ks)$ could also affect the $\Bs2mumu$ decay.
In this case NP enters via a new Wilson coefficient $C_{10}^\NP$ in the branching ratio
\begin{equation}
\Br(B_s\to\mu^+\mu^-)
= \Br(B_s\to\mu^+\mu^-)_\SM \left|1
	+\frac{C_{10}^{\mu\NP}}{C_{10}^{\mu\SM}}\right|^2~,
\label{Br_th}
\end{equation}
where $C_{10}^{\mu\SM}$ is the SM value of $C_{10}^\mu$,
$C_{10}^\SM =-4.41$ \cite{Damir1205,DAlise2403}. 
\par
Now we parameterize the Wilson coefficient as mentioned in Sec.\ I, 
\begin{equation}
C_{9,10}^{\ell,\NP} 
= \calN  A_{9,10}^\ell\left(\frac{v}{M_{\NP}}\right)^\alpha ~,
\label{C_setup}
\end{equation}
where $\ell=e,\mu$ and $\calN=|\alpha_{em}V_{tb}V_{ts}^*|^{-1}$.
Here $M_\NP$ is the NP scale and $A_{9,10}^\ell$ are the involved coefficients, and
$\alpha$ is a free parameter.
$A_{9,10}^\ell$ involve the fermionic couplings to NP.
%
%
%
%
%
In general NP effects could exist in the electron sector with $A_{9,10}^e\ne 0$.
But non-zero $A_{9,10}^e$ would affect the electron $(g-2)$ which is in  good 
agreement with the SM.
For example, in \cite{Cheung2204} the leptoquark couplings $X_{31}^{LR}$ can contribute both
to the electron anomalous moment and $R(\Ks)$.
So we simply assume that the NP appears only in the muon sector putting $A_{9,10}^e=0$,
and we drop the superscript $\ell$ or $\mu$ in what follows.
\par
%
%
%
%
%
Our parametrization encodes explicitly the heavy, tree-level NP mediators including unparticles.
And it is well known that unparticles can be described by an infinite tower of massless particles
\cite{Stephanov0705, JPL0901}, 
so the Lagrangian of that kind can also be covered by Eq.\ (\ref{C_setup}).
When low-energy effective field theory (LEFT) is matched onto 
standard model effective field theory (SMEFT), one can find a similar form with $\alpha=2$
(see, e,g., \cite{Chen2401}).
One merit of Eq.\ (\ref{C_setup}) is that $M_\NP$ appears explicitly, 
so one can estimate the NP scale more generically.
But if the NP Lagrangian contains light NP elements which contribute to the $B^+\to K^+ +$ missing energy,
it cannot be described by our framework.
Light NP such as sterile neutrinos or dark fermions can possibly explain $\Br(\B2Knunu)$,
which is however beyond the scope of current analysis.
\par
In principle it is possible that NP could appear in $C_7$.
But as discussed in \cite{Bardhan2107,JPL2208}, 
the constraint on $C_7$ from $B\to X_s\gamma$ is so strong that there is very little room for NP in $C_7$,
so we neglect possible NP in $C_7$.
%
%
%
\par
Now let's move to $b\to s\nu\nu$ decay.
The relevant effective Hamiltonian is \cite{Allwicher2309,Sumensari2406}
\begin{equation}
\calH_{\rm eff}(b\to s\nu\nu) =
\frac{4G_F}{\sqrt{2}}\frac{e^2}{(4\pi)^2}V_{tb}V_{ts}^*
\sum_{k=L,R}\sum_{i,j}C_{\nu k}^{ij}\Big[\sbar\gamma_\mu P_k b\Big]
    \Big[{\bar\nu}_i\gamma^\mu(1-\gamma_5)\nu_j\Big] + {\rm h.c.}~.
\end{equation}

The branching ratio of $B^+\to K^+ \nu\nu$ is \cite{Becirevic2301,Allwicher2309}
\begin{equation}
\Br(B^+\to K^+ \nu\nu)=
\Br(B^+\to K^+ \nu\nu)_\SM\frac{1}{3|C_{\nu L}^\SM|^2}\sum_{i,j}\Big|C_{\nu L}^{ij}+C_{\nu R}^{ij}\Big|^2~,
\label{Brnu}
\end{equation} 
where $C_{\nu L}^{ij}=C_{\nu L}^\SM\delta^{ij}+\delta C_{\nu L}^{ij}$ and
$C_{\nu L}^\SM$ is the SM Wilson coefficient, $C_{\nu L}^\SM =-6.32$ \cite{Sumensari2406}.
The NP effects appear through $\delta C_{\nu L}^{ij}$ and $C_{\nu R}^{ij}$.
We can expand the branching ratio as follows:
\begin{equation}
\Br(B^+\to K^+ \nu\nu)=
\Br(B^+\to K^+ \nu\nu)_\SM\Big|1+L+Q\Big|~,
\label{Brnu2}
\end{equation}
where linear(L) and quadratic(Q) terms of $C_\nu$s are parameterized as
\begin{eqnarray}
L&=&  \frac{\calN}{C_L^\SM}
           A_L\left(\frac{v}{M_\NP}\right)^\alpha~,\nn\\
Q&=&  \left|\frac{\calN}{C_L^\SM}\right|^2
           A_Q\left(\frac{v}{M_\NP}\right)^{2\alpha}~.
\label{LnQ}
\end{eqnarray}
%
%
%
%
%
The idea of this parametrization is basically the same as that of Eq.\ (\ref{C_setup}).
%
%
%
%
%
\section{Results and discussions}
%
%
%
We implement the $\chi^2$ fit for $R(\Ks)$ and the branching ratios $\Br(\Bs2mumu)$ and $\Br(\B2Knunu)$.
The scan range covers $0\le\alpha\le 5$, $1~\TeV\le M_\NP\le 50~\TeV$, 
$|A_{9,10}|\le 1$, $|A_L|\le 10$, and $|A_Q|\le 10^2$.
The best-fit values are summarized in Table \ref{T_best_fit}.
\begin{table}
\begin{tabular}{cc|cc}\hline\hline
$\alpha$                   & $~~~1.53~~~~$      & $R(K)_L$     & $0.971$     \\
$M_{\rm NP}$ (TeV) & $~~~26.4~~~$     & $R(K)_C$     & $0.978$    \\
$~~~A_9~~~$          &  $~~~0.064~~~$    & $R(K^*)_L$   & $0.918$    \\
$~~~A_{10}~~~$     &  $~~~0.115~~~$    & $R(K^*)_C$   & $0.967$     \\
$~~~A_L\times 10~~~$   &  $~~~0.791~~~~$    & $~~~\Br(B_s\to\mu^+\mu^-)$  & $~~~~~3.36\times 10^{-9}$   \\
$~~~A_Q\times 10^2~~~$  &   $~~~ -0.722~~~$   & $~~~\Br(B^+\to K^+\nu\nu)$  & $~~~~~2.29\times 10^{-5}$ \\
$~~~C_9^\NP~~~$   & $~~~0.099~~~$           &  $C_{10}^\NP$    &  $0.176$ 
\\ \hline\hline
\end{tabular}
\caption{Best-fit values of our fitting.}
\label{T_best_fit}
\end{table}
%
The minimum value of $\chi^2$ per d.o.f is $\chi^2_{\rm min}/{\rm d.o.f}=1.37$.
The best-fit $\alpha_\best$ is less than 2.
The best-fit $M_{\NP,\best}$ is 26.4 TeV.
If $\alpha_\best$ were larger then $M_{\NP,\best}$ would be smaller.
\par
\begin{figure}
\begin{tabular}{cc}
\hspace{-1cm}\includegraphics[scale=0.12]{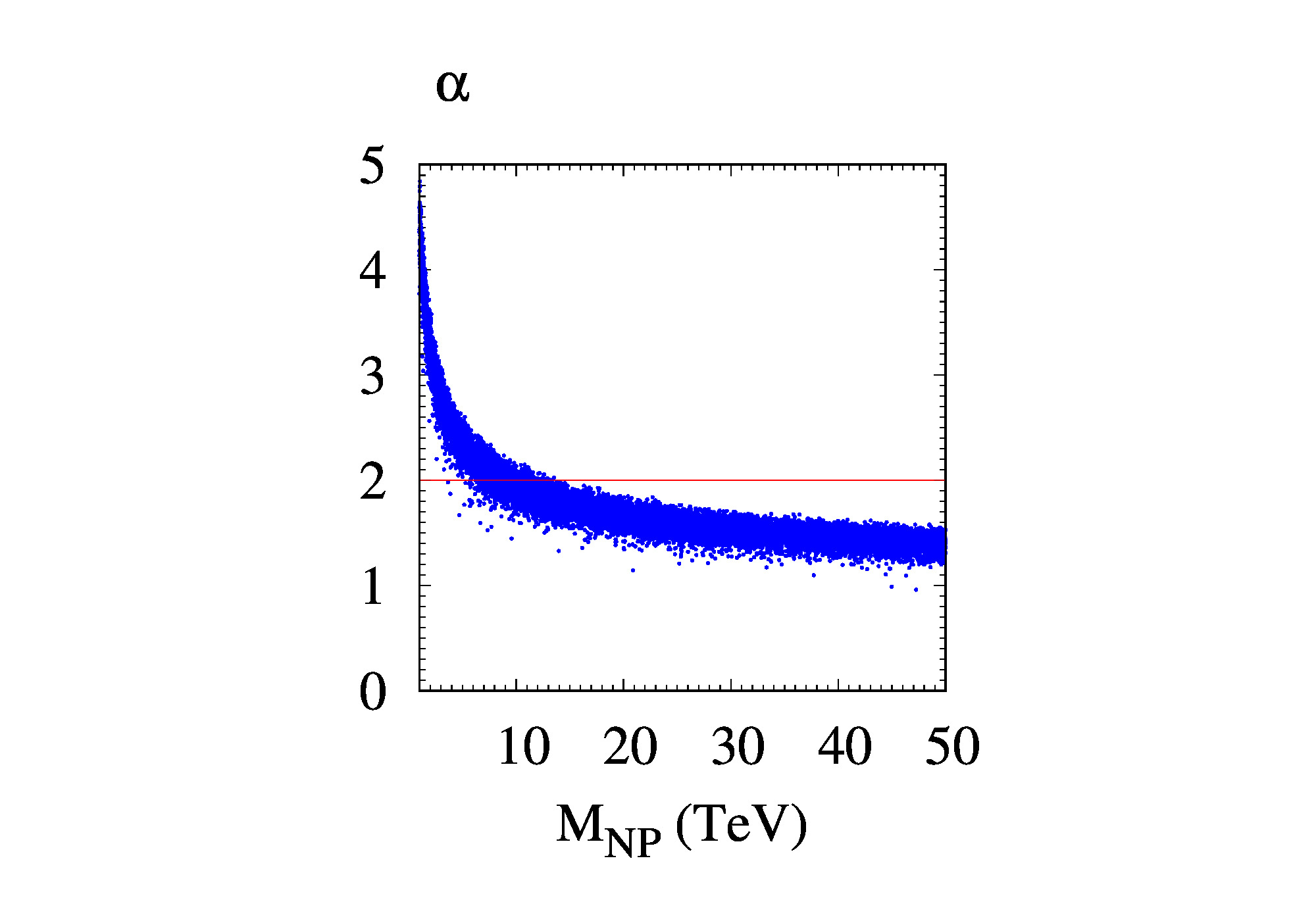} &
\hspace{-1cm}\includegraphics[scale=0.12]{A10-A9.jpg} \\
(a) & (b) \\
\hspace{-1cm}\includegraphics[scale=0.12]{C10-C9.jpg} &
\hspace{-1cm}\includegraphics[scale=0.12]{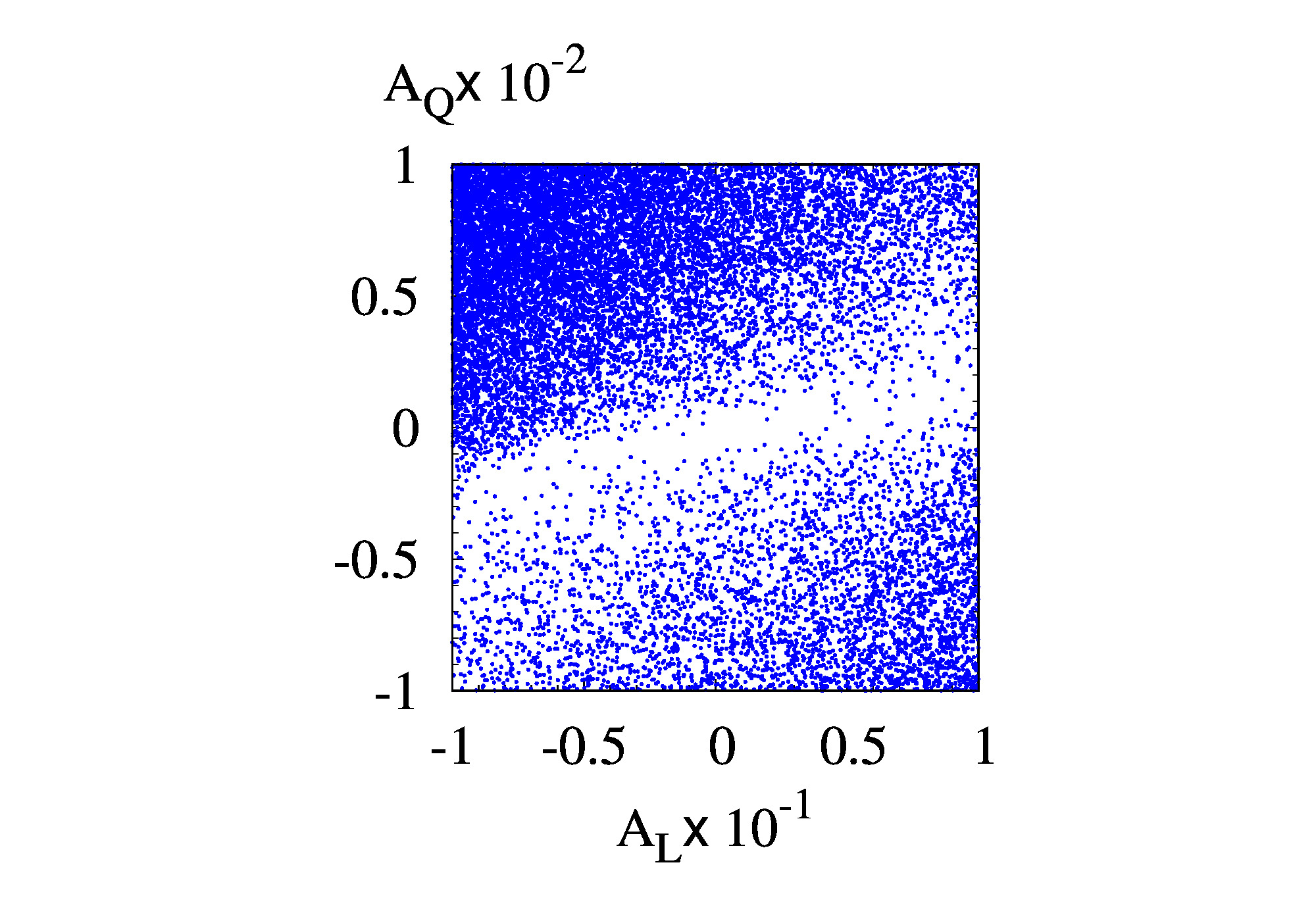} \\
(c) & (d) \\
\hspace{-1cm}\includegraphics[scale=0.12]{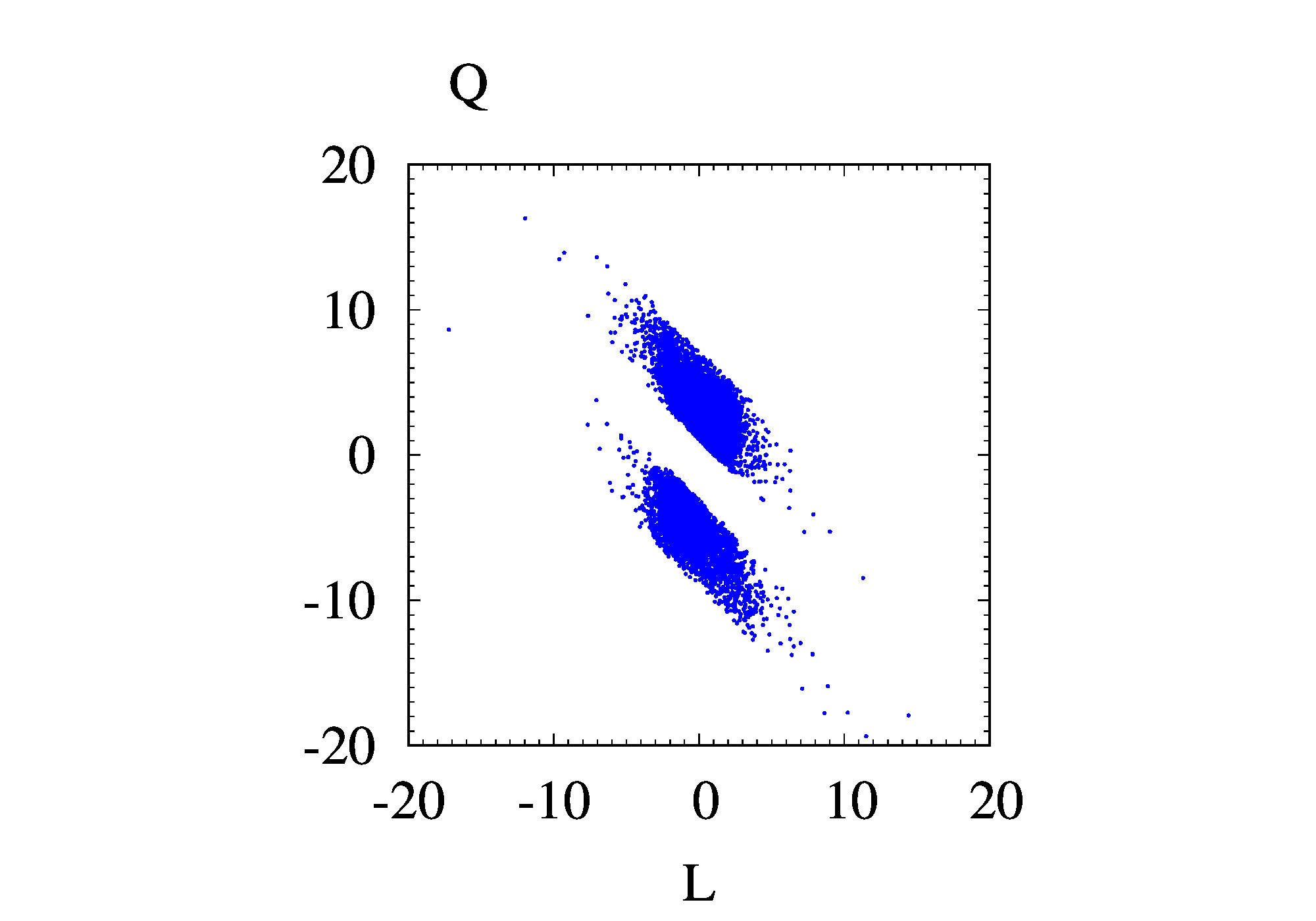} & \\
(e) &
\end{tabular}
\caption{\label{fit_para} Allowed regions of 
(a) $\alpha$ vs. $M_\NP$, 
(b) $A_{10}$ vs. $A_9$,
(c) $C_{10}^\NP$ vs. $C_9^\NP$, 
(d) $A_Q$ vs. $A_L$, and
(e) $Q$ vs. $L$ at the $2\sigma$ level.
}
\end{figure}

Figure \ref{fit_para} shows the allowed regions of the fitting parameters,
Wilson coefficients, $L$, and $Q$ at the $2\sigma$ level.
In Fig.\ \ref{fit_para} (a), $\alpha$ and $M_\NP$ show a typical exponential behavior.
One point to note is that for those vales of $\alpha\gtrsim 1.6$ the allowed $M_\NP$ has
an upper bound.
This is mainly because $\Br(B^+\to K^+\nu\nu)_\exp$ is far away from the SM calculations.
If NP could explain the discrepancy, its effect should be large enough.
To make it happen for moderate values of the relevant couplings, 
$M_\NP$ must be bounded from above.
If we do not consider $\B2Knunu$ process, then one can find no upper bounds of $M_\NP$.
\par
As shown in Fig. \ref{fit_para} (b), $A_{10}$ favors positive values. 
It reflects the fact that the experimental data of $\Br(B_s\to\mu\mu)$ are 
slightly smaller than the SM predictions.
Note that $C_{10}^\SM$ in $\Br(\Bs2mumu)$ is negative.
In Fig.\ \ref{fit_para} (c) there are very few points around $C_{10}^\NP\sim 8.5$ which are not shown here.
Those few points are possible because the branching ratio $\Br(B_s\to\mu^+\mu^-)$ is 
proportional to $\sim\Big|1+C_{10}^\NP/C_{10}^\SM\Big|^2$, and we have two solutions of
positive and negative $\Big(1+C_{10}^\NP/C_{10}^\SM\Big)$.
Large $C_{10}^\NP$ corresponds to the negative solution (note that $C_{10}^\SM<0$).
See Figs.\ 2 (c) and (d) of \cite{JPL2208}.
In Figs.\ \ref{fit_para} (d) and (e) there are narrow strips of rare points which are not allowed by our fitting.
They correspond to the line of $L+Q=0$ in Eq.\ (\ref{Brnu2}) for which NP effects vanish.
\par
\begin{figure}
\begin{tabular}{cc}
\hspace{-1cm}\includegraphics[scale=0.12]{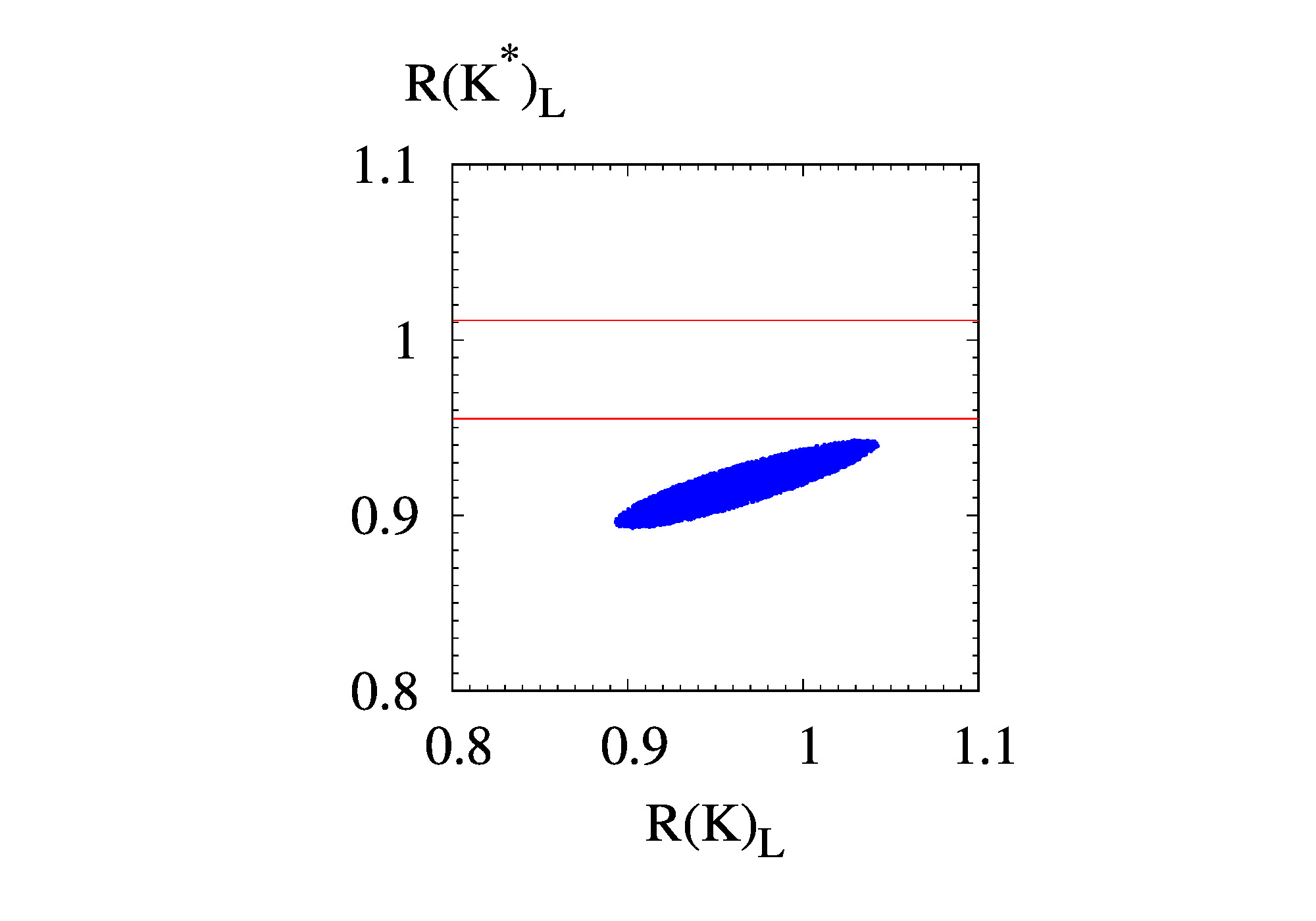} &
\hspace{-1cm}\includegraphics[scale=0.12]{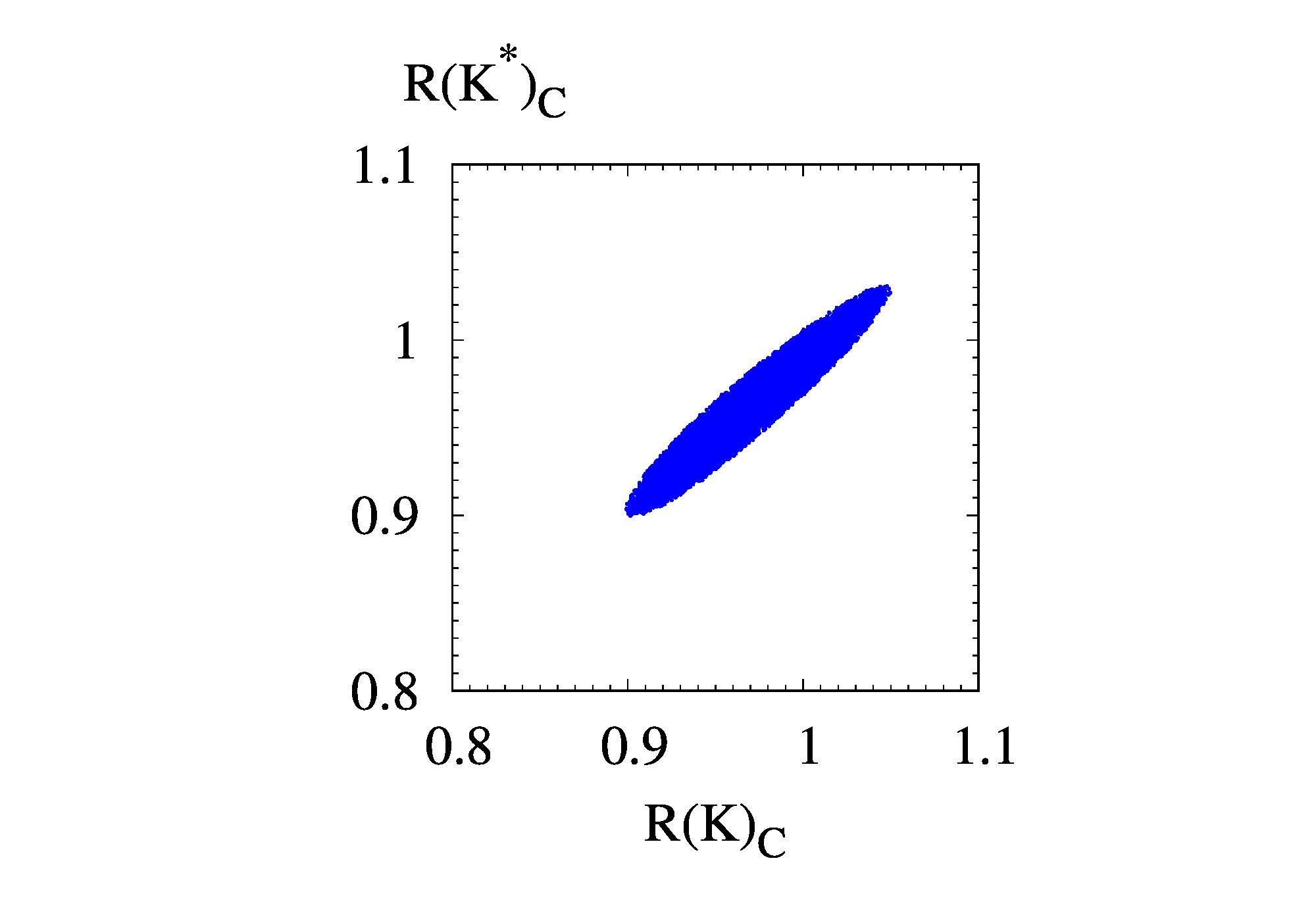} \\
(a) & (b) \\
\hspace{-1cm}\includegraphics[scale=0.12]{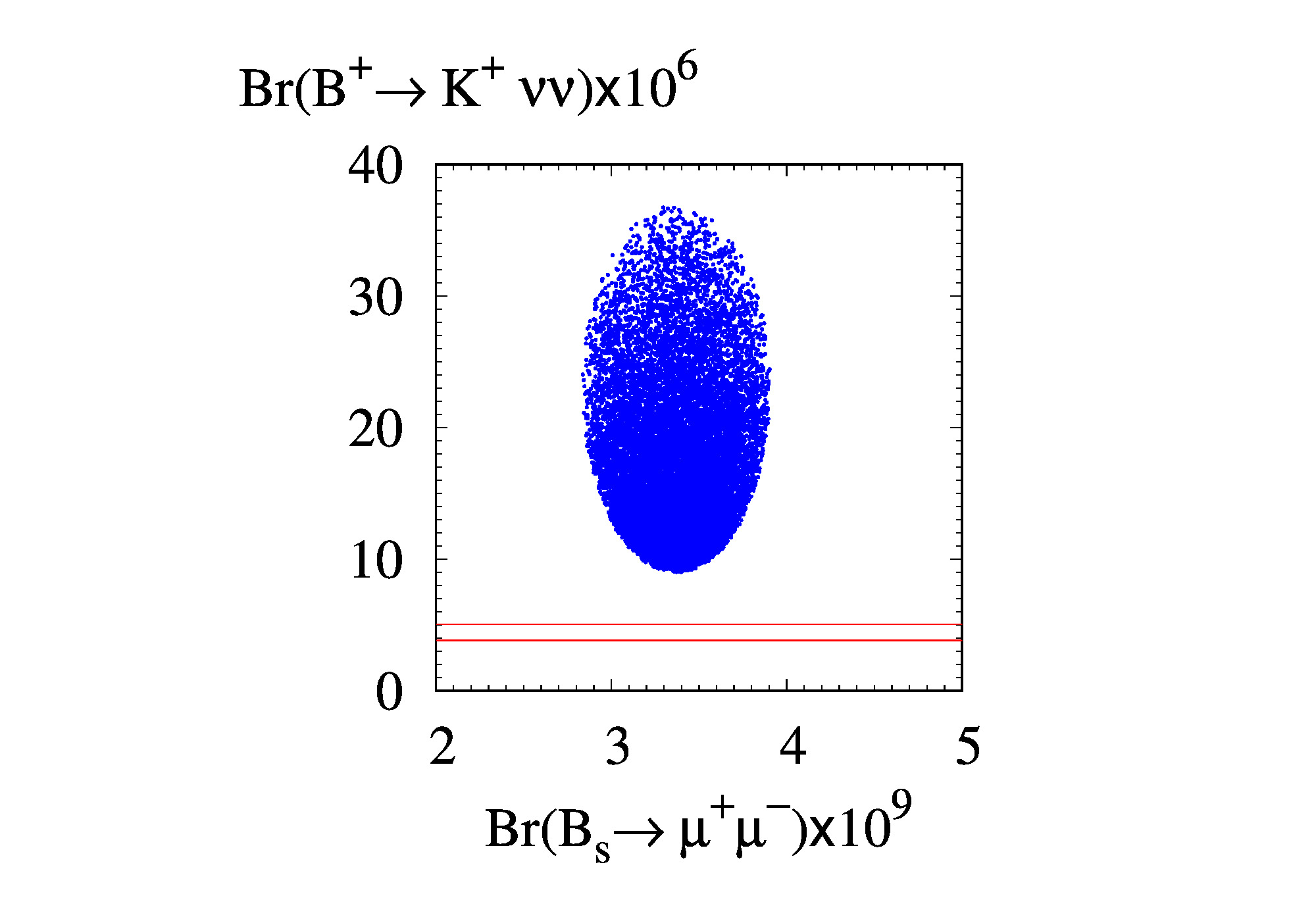} & \\
(c) &
\end{tabular}
\caption{\label{RKsBr} Allowed regions of
(a) $R(K^*)_L$ vs. $R(K)_L$,
(b) $R(K^*)_C$ vs. $R(K)_C$, and
(c) $\Br(B^+\to K^+\nu\nu)$ vs. $\Br(B_s\to\mu^+\mu^-)$ at the $2\sigma$ level.
Horizontal lines show the SM ranges in $2\sigma$ 
of $R(K^*)_L$ in (a) and of $\Br(B^+\to K^+\nu\nu)$ in (c).
}
\end{figure}

In Fig.\ \ref{RKsBr}, we plot the allowed regions of the observables.
As shown in Fig.\ \ref{RKsBr} (a) and (c), in our fitting $R(K^*)_L$ and $\Br(B^+\to K^+\nu\nu)$ have
no overlaps with the SM predictions while $R(K)_L$, $R(\Ks)_C$ and
$\Br(\Bs2mumu)$ are quite compatible with SM.
While the experimental data of $R(K^*)_L$ in Eq.\ (\ref{RKsLHCb}) are compatible with the SM,
the central value of it is rather smaller than the SM compared to other data.
%
%
%
%
\par
\begin{figure}
\begin{tabular}{cc}
\hspace{-1cm}\includegraphics[scale=0.12]{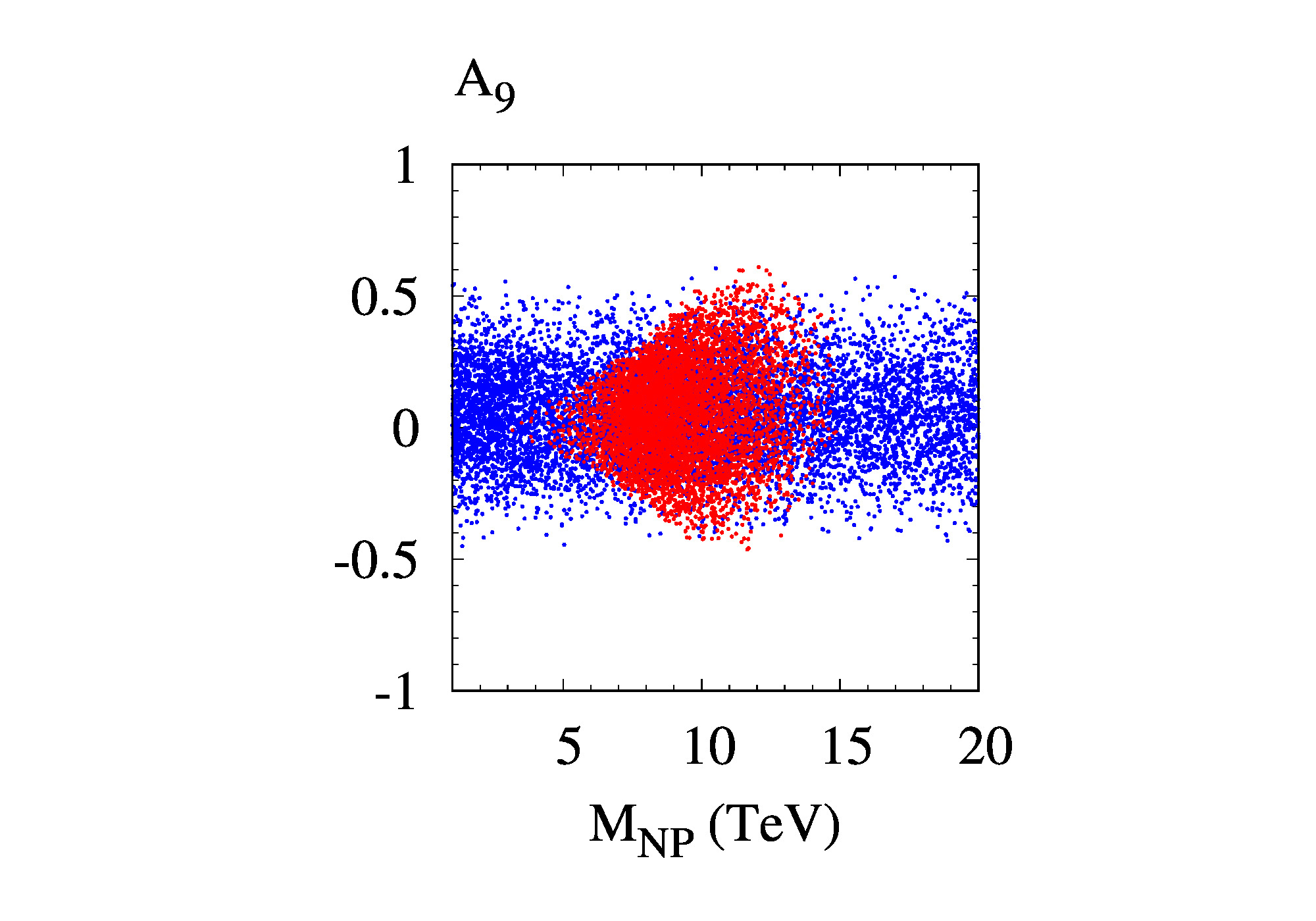}&
\hspace{-1cm}\includegraphics[scale=0.12]{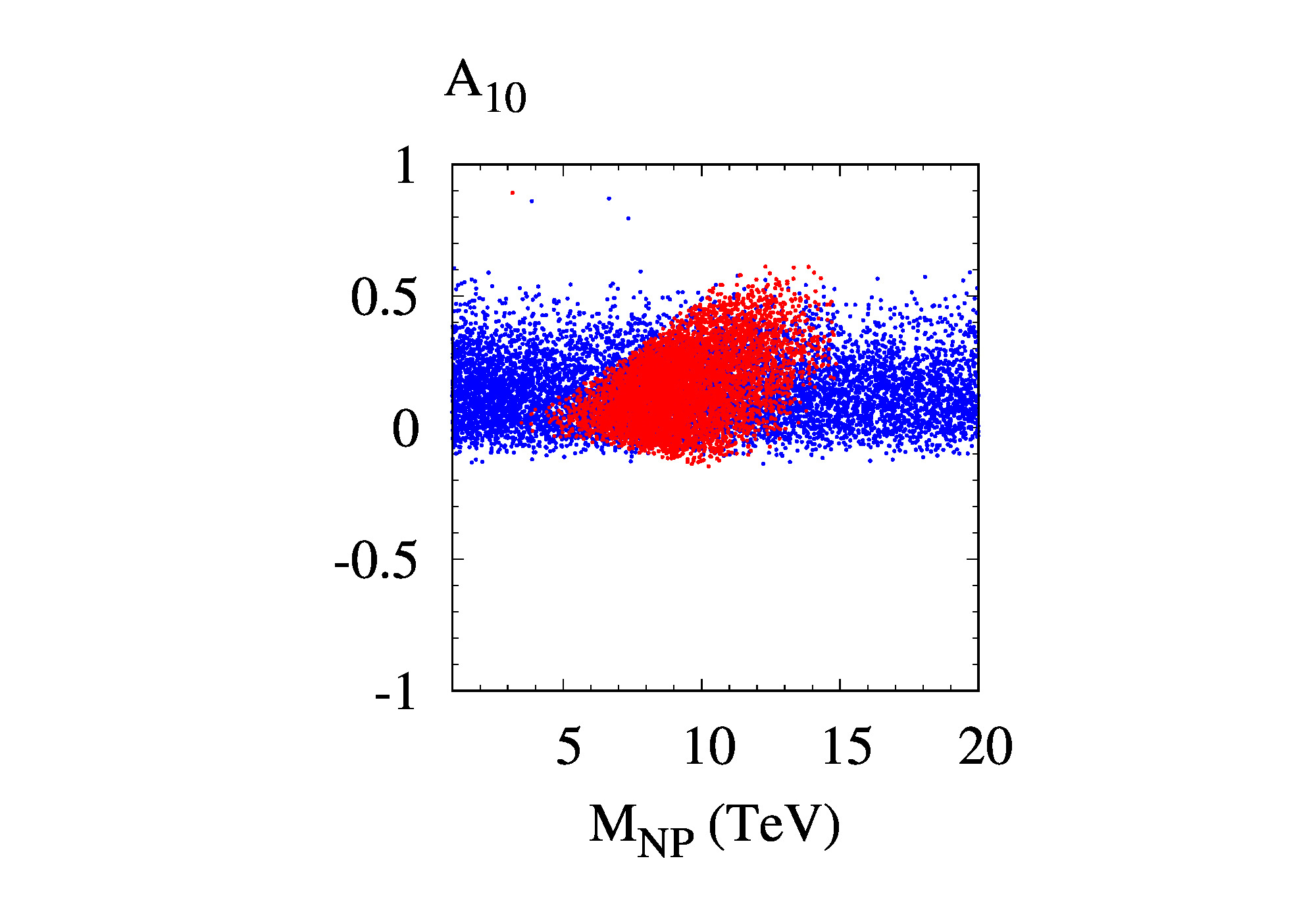}\\
(a) & (b) \\
\hspace{-1cm}\includegraphics[scale=0.12]{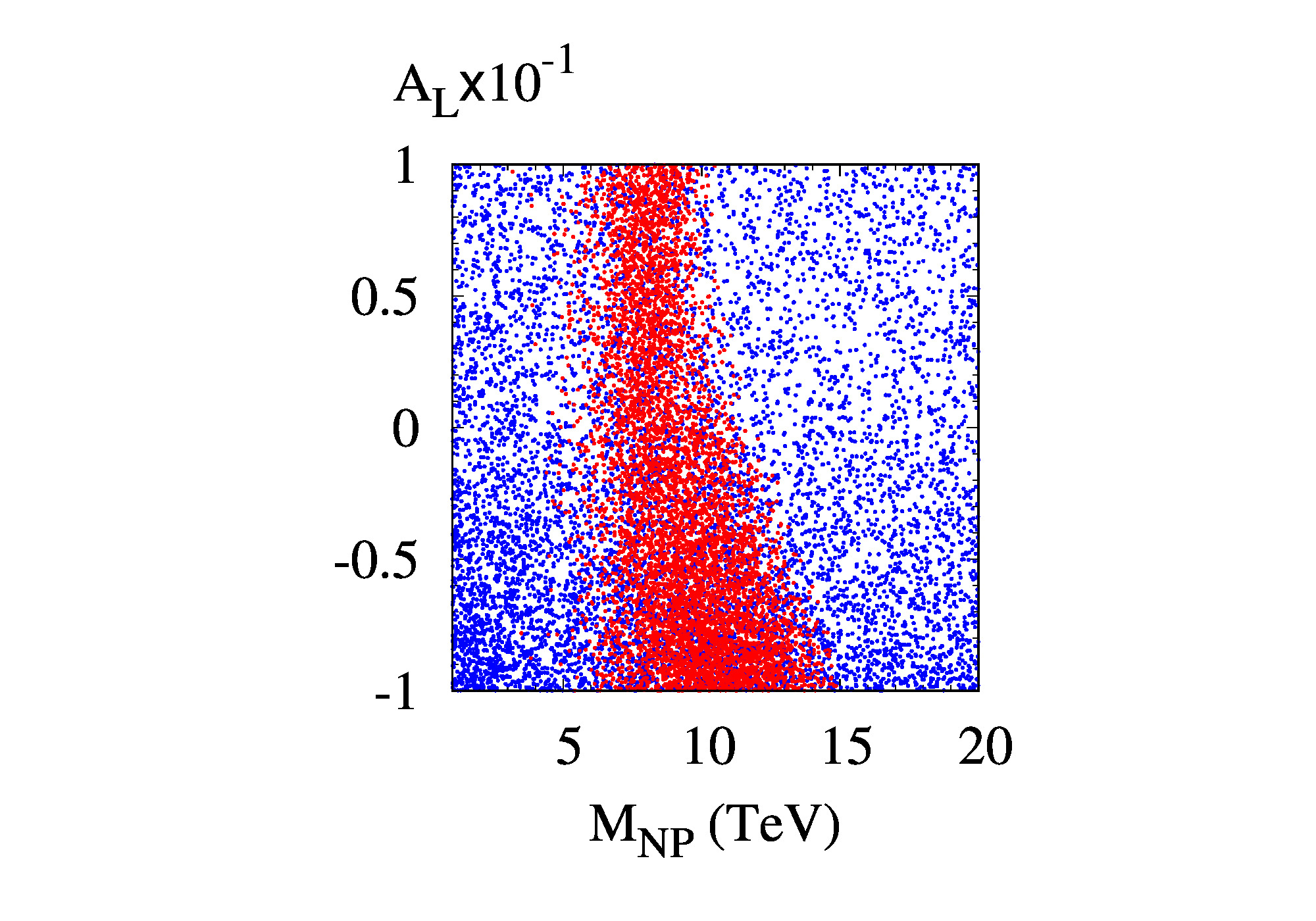}&
\hspace{-1cm}\includegraphics[scale=0.12]{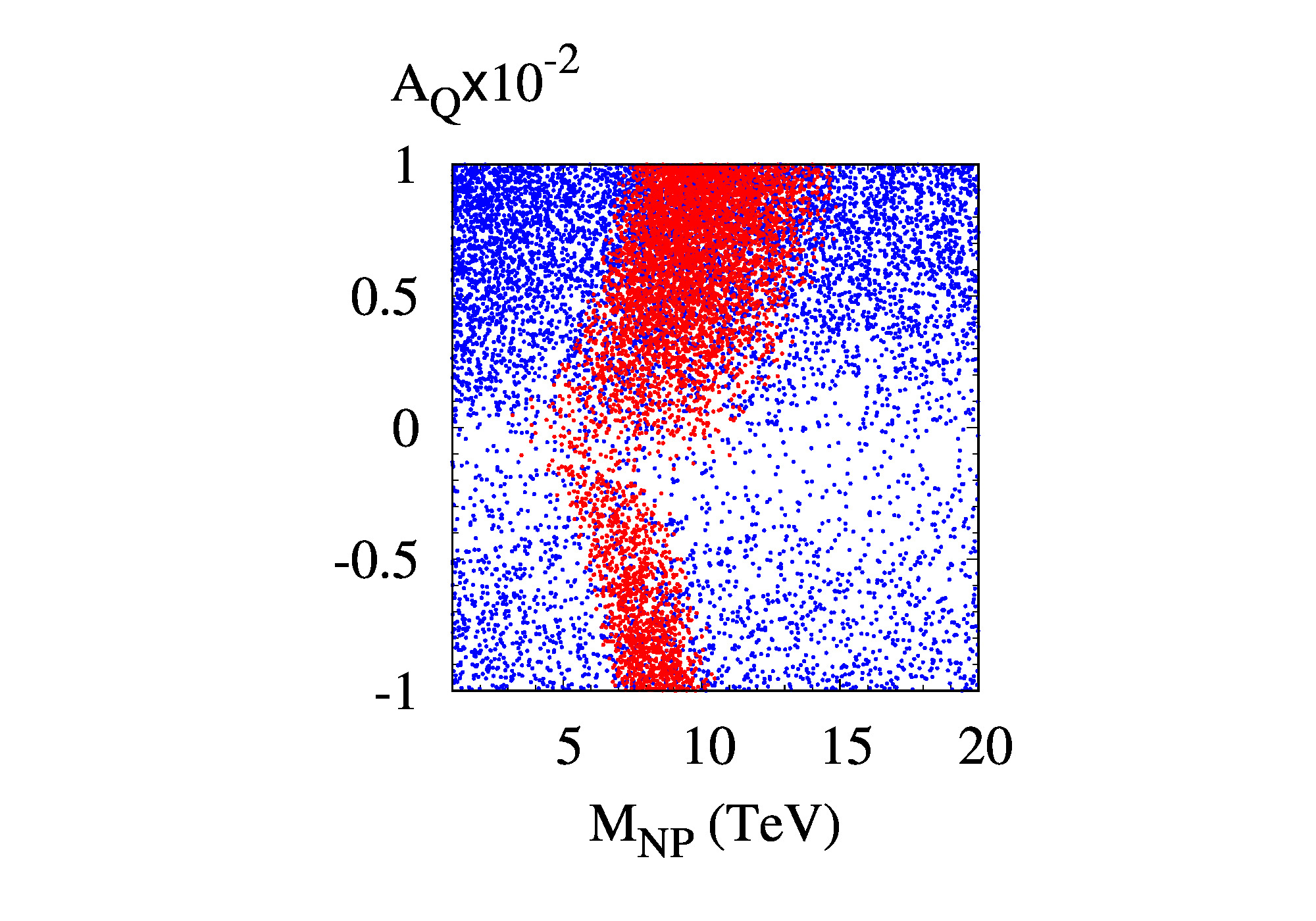}\\
(c) & (d) \\
\hspace{-1cm}\includegraphics[scale=0.12]{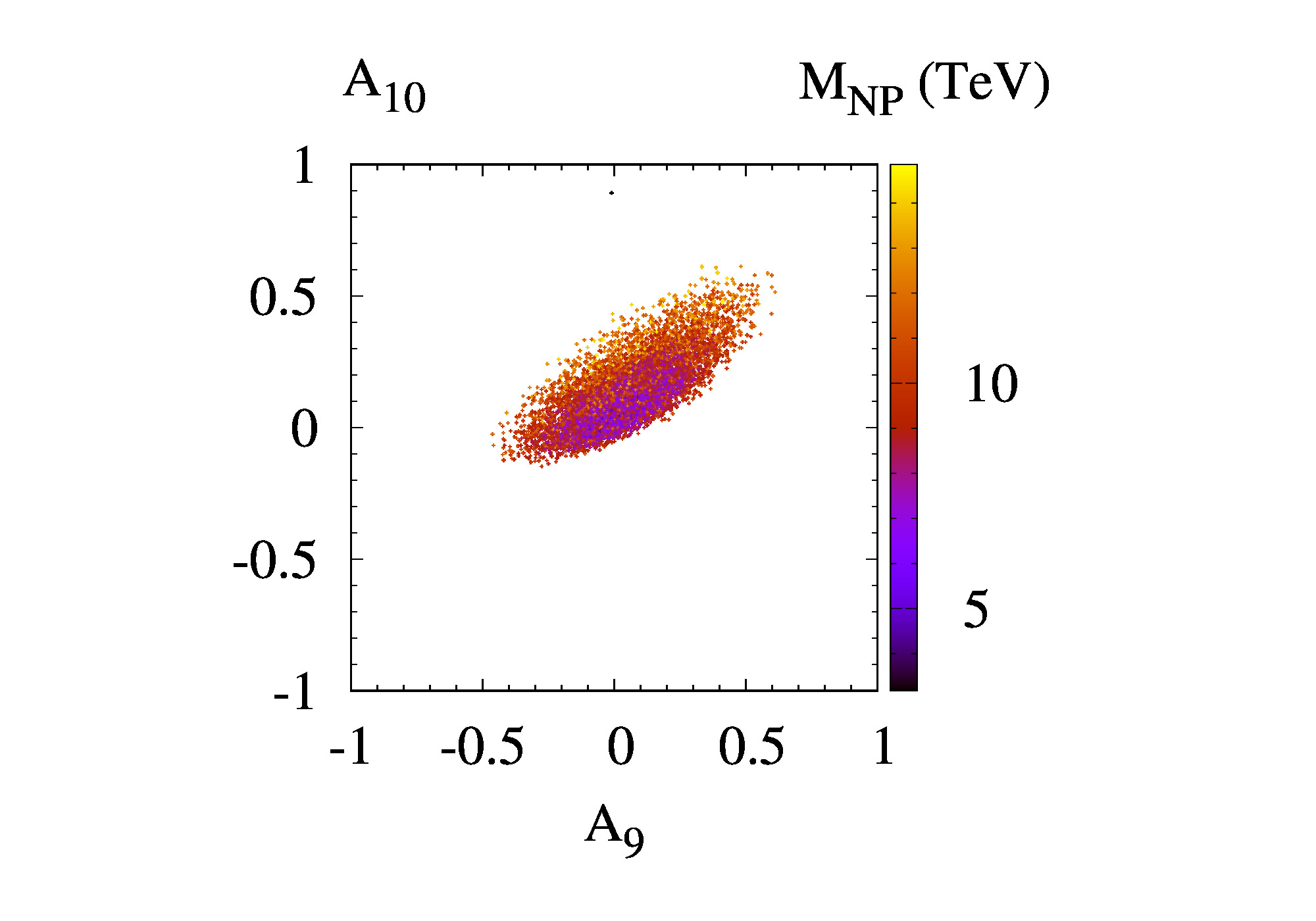}&
\hspace{-1cm}\includegraphics[scale=0.12]{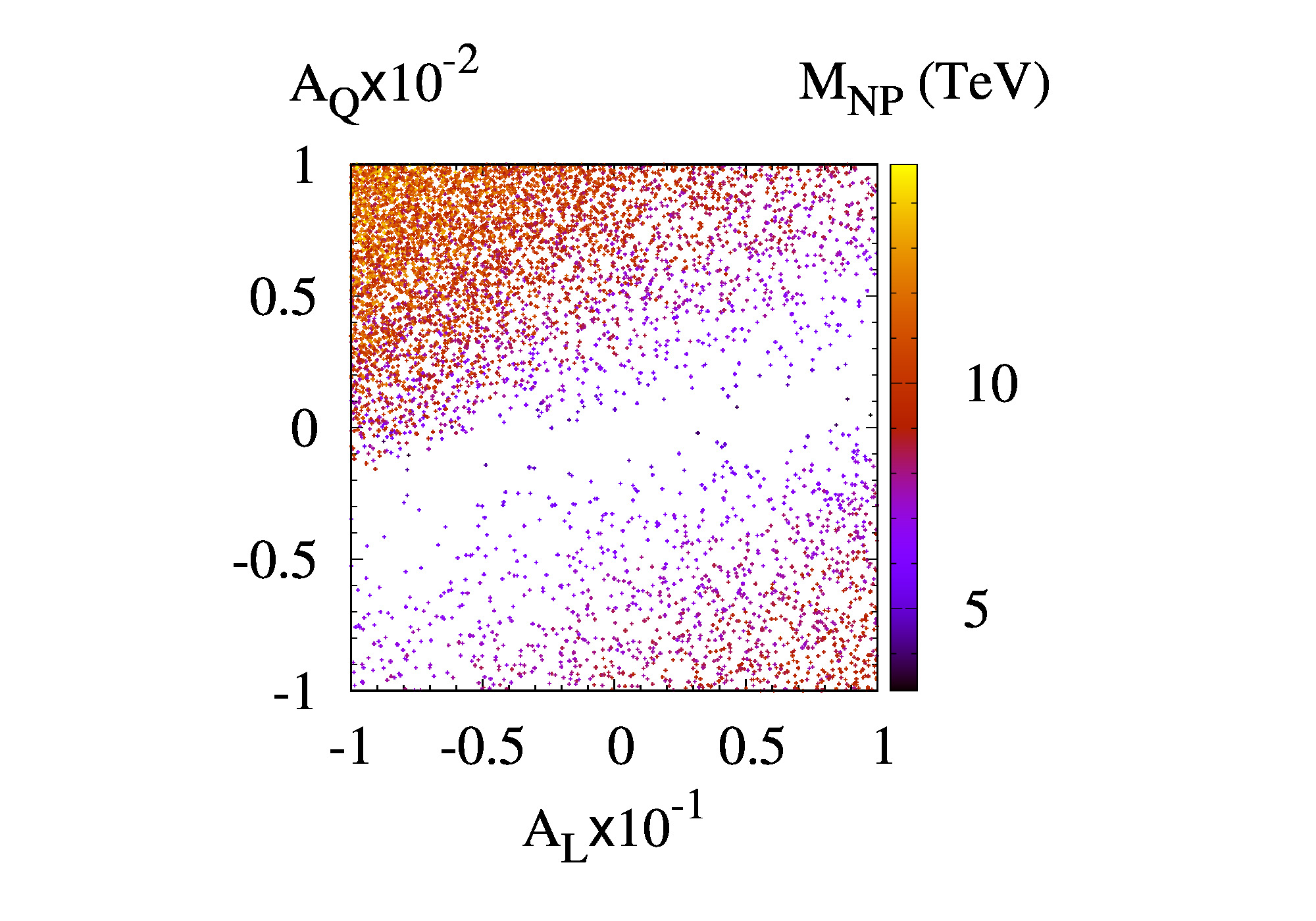}\\
(e) & (f)
\end{tabular}
\caption{\label{Aj_alp2} Allowed regions (at the $2\sigma$ level) of
(a) $A_9$ vs. $M_\NP$,
(b) $A_{10}$ vs. $M_\NP$,
(c) $A_L$ vs. $M_\NP$, and
(d) $A_Q$ vs. $M_\NP$
for free $\alpha$ (blue) and fixed $\alpha=2$ (red) respectively; 
(e) $A_{10}$ vs. $A_9$ with respect to $M_\NP$, and
(f) $A_Q$ vs. $A_L$ with respect to $M_\NP$ for fixed $\alpha=2$. 
}
\end{figure}

Figure \ref{Aj_alp2} provides $A_j$ vs. $M_\NP$ for fixed $\alpha=2$.
This is the case for ordinary new particle contributions.
Note that in case of $\alpha=2$, $M_\NP$ is restricted from above and below.
This is expected in Fig.\ \ref{fit_para} (a). 
For $\alpha=2$, we have 
\begin{equation}
3.17~\TeV\le M_\NP\le 14.9~\TeV~.
\label{MNPwindow}
\end{equation}
Comparing Figs.\ \ref{Aj_alp2} (a) and (b) with (c) and (d), 
$M_\NP$ is more sensitive to $A_{L,Q}$ in the sense that
for larger $|A_{L,Q}|$, the upper bound of $M_\NP$ also gets larger.
The feature can also be seen in Figs.\ \ref{Aj_alp2} (e) and (f).
As for $A_{9,10}$, they are responsible for $R(\Ks)$ and $\Br(\Bs2mumu)$ 
which are in agreement with the SM.
Thus $A_{9,10}$ are much more involved in the lower bound of $M_\NP$ for which 
$A_{9,10}$ get close to zero as shown in Figs.\ \ref{Aj_alp2} (a) and (b).
On the contrary, $A_{L,Q}$ are responsible for $\Br(\B2Knunu)$ that deviates from the SM,
which makes $A_{L,Q}$ much more involved with the upper bound of $M_\NP$.
As for $\Br(B^+\to K^+\nu\nu)$ finite NP effects should be maintained to accommodate
the experimental data, so for given $A_{L,Q}$, $M_\NP$ could not be large enough.
This is the origin of the upper bound of $M_\NP$.
According to Figs.\ \ref{Aj_alp2} (c), (d) and (f), negative $A_L$ and positive $A_Q$ are
more involved with the upper bound of $M_\NP$.
These parts tend to enhance $\Br(\B2Knunu)$ in Eq.\ (\ref{Brnu2}) to come up with the
experimental data.
%
%
\par
It is quite remarkable that our generic analysis puts a window of allowed $M_\NP$ 
with the width of $\sim 12~\TeV$.
Using the window one can check the validity of various NP scenarios with their couplings and NP scales
which are constrained by other experimental data.
\par
\begin{figure}
\begin{tabular}{cc}
\hspace{-1cm}\includegraphics[scale=0.12]{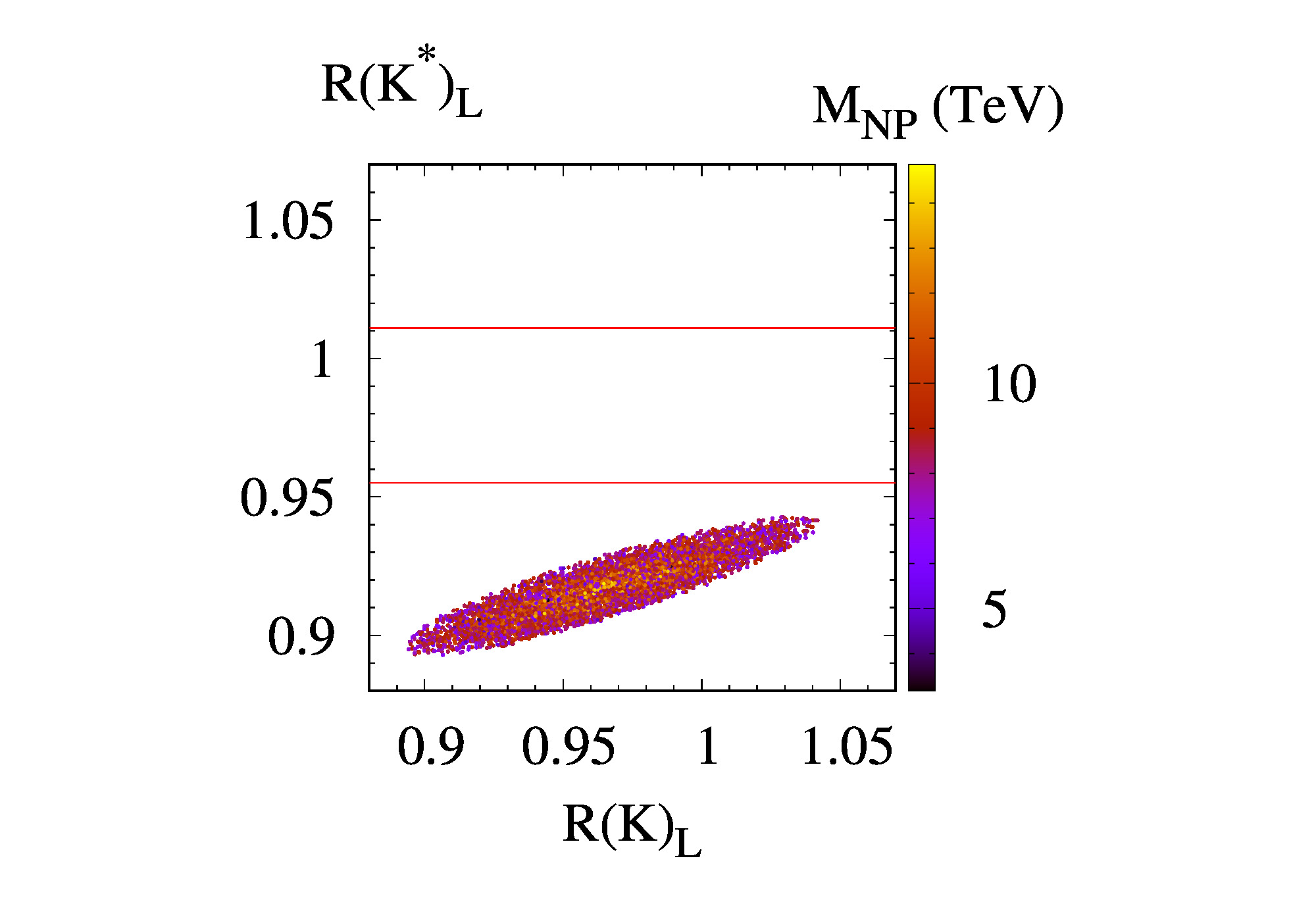} &
\hspace{-1cm}\includegraphics[scale=0.12]{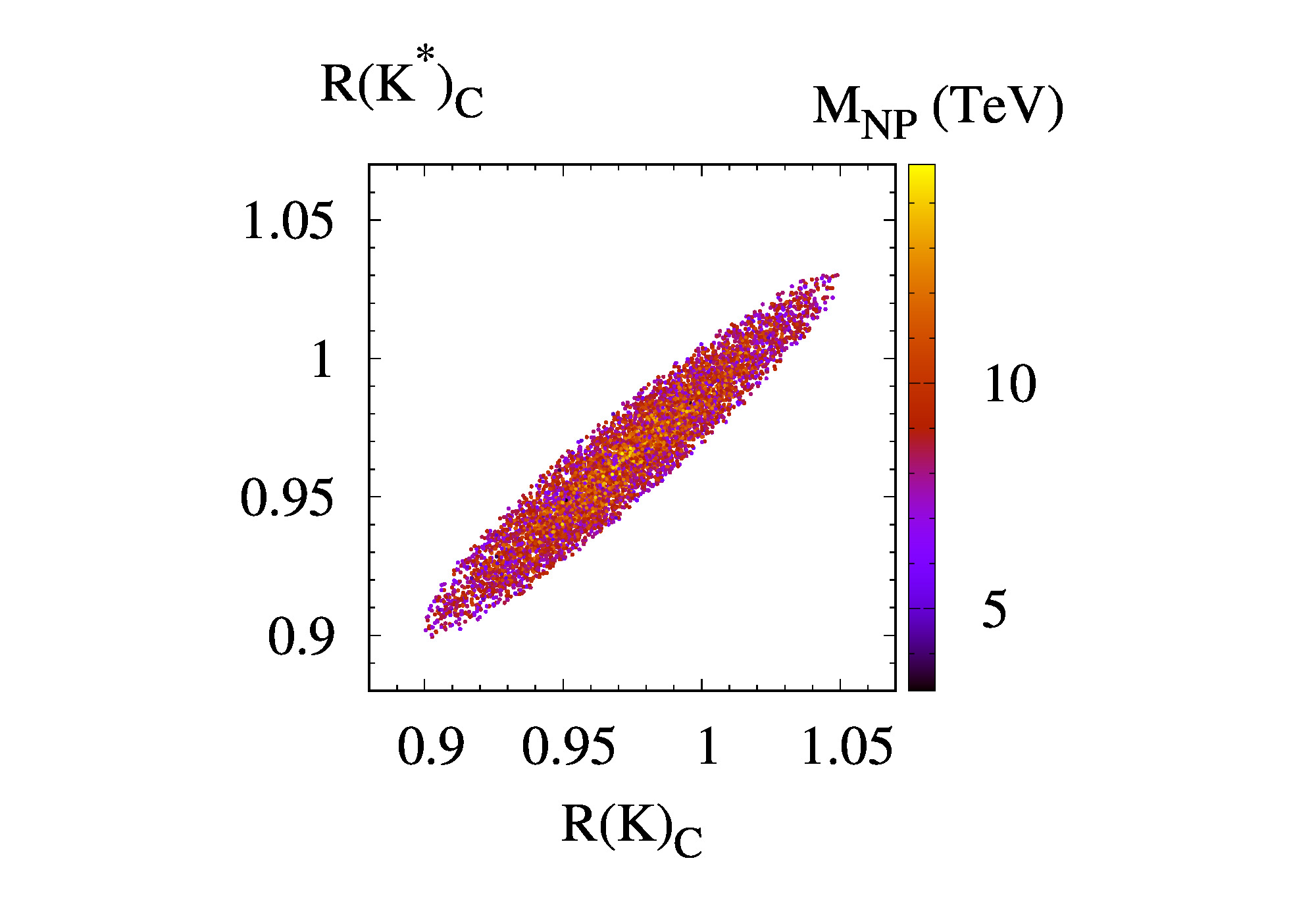} \\
(a) & (b) \\
\hspace{-1cm}\includegraphics[scale=0.12]{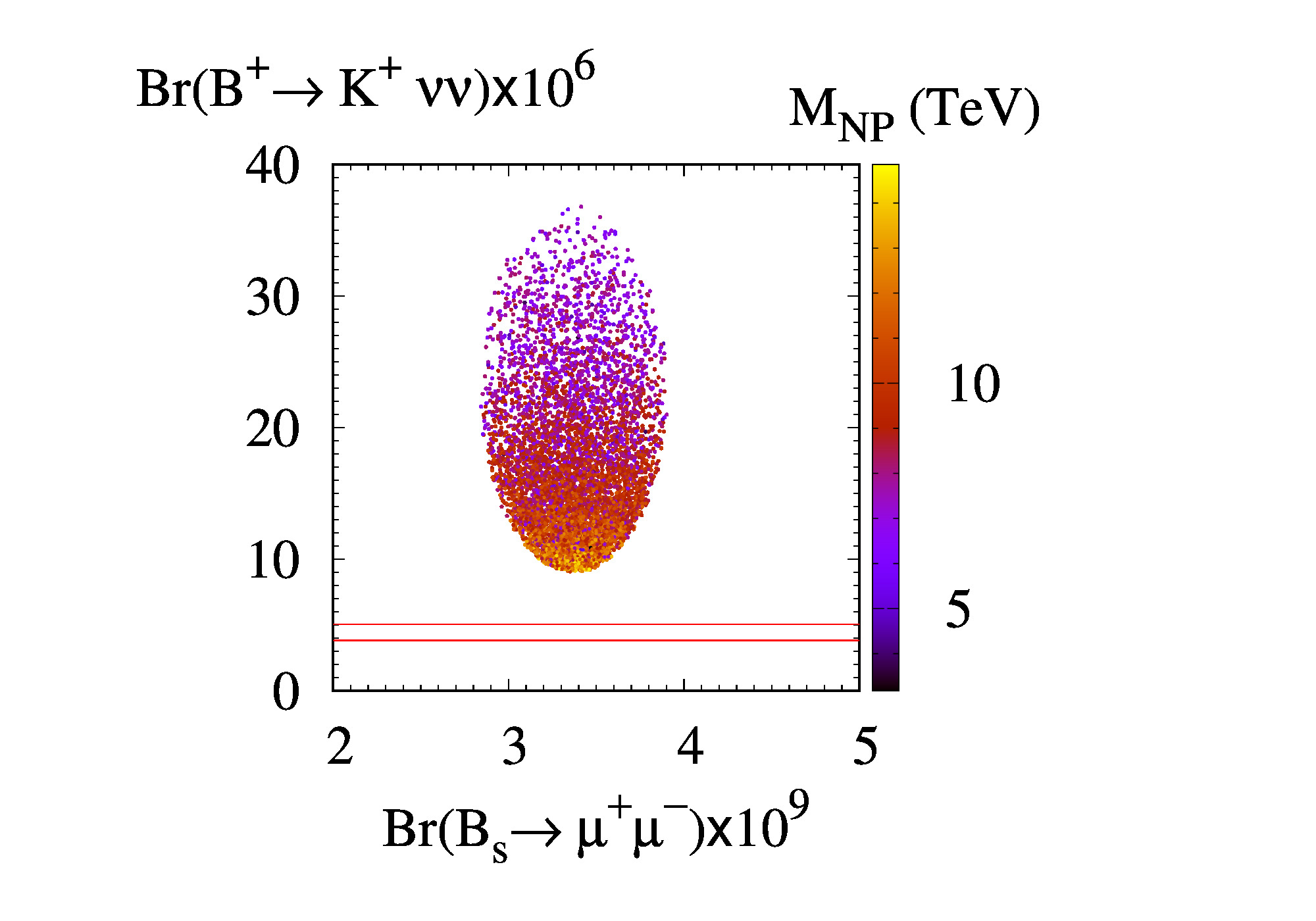} & \\
(c) &
\end{tabular}
\caption{\label{3G_alp2} Allowed regions of
(a) $R(K^*)_L$ vs. $R(K)_L$ with respect to $M_\NP$,
(b) $R(K^*)_C$ vs. $R(K)_C$ with respect to $M_\NP$, and
(c) $\Br(B^+\to K^+\nu\nu)$ vs. $\Br(B_s\to\mu^+\mu^-)$ with respect to $M_\NP$
at the $2\sigma$ level for fixed $\alpha=2$.
Horizontal lines show the SM ranges in $2\sigma$ 
of $R(K^*)_L$ in (a) and of $\Br(B^+\to K^+\nu\nu)$ in (c).
}
\end{figure}

Figure \ref{3G_alp2} plots the allowed regions of observables with respect to $M_\NP$ for fixed $\alpha=2$.
For larger values of $M_\NP$, allowed regions of $R(\Ks)$ and the branching ratios of
$\Bs2mumu$ and $\B2Knunu$ get smaller.
This is because the NP effects tend to diminish for larger $M_\NP$.
Compared with Fig.\ \ref{RKsBr}, fixing $\alpha=2$ has almost no effects on the
allowed regions of $R(\Ks)_{L,C}$ and the branching ratios $\Br(B_s\to\mu^+\mu^-)$ 
and $\Br(B^+\to K^+\nu\nu)$.
It means that although $\alpha$ is fixed to 2, other parameters can cover the observables 
but the allowed regions of them are shrunk much, as shown in Fig.\ \ref{Aj_alp2}.
%
%
%
%
\par
Our approach can be used as a toolkit to test the validity of specific NP models.
For example, 
allowed $Z'$ mass for explaining $(g-2)_\mu$ from $U(1)_{B_2-L_\mu}$ model in \cite{Cen2104} is
$m_{Z'}< 300~\GeV$ for a coupling of unity. 
This is off from our suggested range of $M_\NP$.
On the other hand, a vector LQ of \cite{Du2104} compatible with $(g-2)_\mu$ and various $B$ anomalies 
including $R(\Ks)$ is estimated to have its mass from 1.5 TeV to several TeV.
The range is quite challenging in view of our window if the lower bound is $~\sim 3.2~\TeV$.
In a similar manner one can check directly the compatibility of other NP models with 
$R(\Ks)$, $\Br(\Bs2mumu)$, and $\Br(\B2Knunu)$.
\par
%
%
%
In experimental prospects, the High-Luminosity LHC (HL-LHC) of $3000~\perfb$ is expected to reach
the discovery potential (exclusion sensitivity) for scalar leptoquark masses ranges from
$1.2$ to $1.7~\TeV$ ($1.4$ to $1.9~\TeV$) \cite{ATL_22_018, CMS_22_001}.
As for a vector leptoquark, the HL-LHC could exclude its mass up to $2.8~\TeV$ \cite{Desai2301}. 
The limits are slightly below our window.
On the other hand, the masses of $Z'$ could be excluded up to $6.5~\TeV$ by the HL-LHC
\cite{ATL_22_018, CMS_22_001}.
According to the window of Eq.\ (\ref{MNPwindow}), it is possible that leptoquarks would not be
observed at the HL-LHC while $Z'$ is promising.
But the best-fit value of $M_\NP=26.4~\TeV$ is out of reach of the $pp$ collision energy of $\sqrt{s}=14~\TeV$ 
at the HL-LHC.
It is possible that the HL-LHC could not probe NP scale directly.
It should be noted that the mass window is dependent on the coupling parameters.
\par
%
%
%
Previous works on $\B2Knunu$ such as 
\cite{Chen2401,Allwicher2309,Bause2309}
suggested the importance of a non-vanishing coupling only to $\tau$ flavor.
The result comes from combined analyses with $B^0\to K^{*0}\nu\nubar$ and/or $R(D^{(*)})$.
Current study embraces for example $C_R^{\nu\tau}$ coupling of \cite{Chen2401} in Eq.\ (\ref{Brnu}),
but more detailed analysis combining  $B^0\to K^{*0}\nu\nubar$, $R(D^{(*)})$, and other
related processes would be needed.
Our analysis puts more focus on $M_\NP$, through which one can check easily validities of specific models.  
In addition, one can probe possible  unparticle-like degrees of freedom.
Since the best-fit value of $\alpha$ is different from 2, 
one should also take into account possibilities other than ordinary heavy particle mediator at tree level .
\par
%
%
%
While experimental data of $R(\Ks)$ and $\Br(\Bs2mumu)$ are consistent with the SM predictions,
there are still other $b\to s\ell^+\ell^-$ processes which are in tension with the SM, such as
$\Br(B^+\to K^+\mu^+\mu^-)$, $\Br(B_s\to\phi\mu^+\mu^-)$, and $\P5p$
\cite{Aebischer1810,Alguero2304,Hurth2310}.
Here $\P5p$ is one of the angular observables of $B^+\to K^{*+}\mu^+\mu^-$ decay,
defined by some combinations of the angular coefficients $J_i^a$ associated with the full differential distribution \cite{Mahmoudi2408}.
In what follows we do our analysis again including $\Br(B^+\to K^+\mu^+\mu^-)$ and $\P5p$.
We do not consider $B_s\to\phi\mu^+\mu^-$ decay because there are still large uncertainties
in the local form factors \cite{Gubernari2206, Capdevila2309}.
%
\begin{table}
\begin{tabular}{cc|cc}\hline\hline
$\alpha$                   & $~~~1.36~~~~$      & $R(K)_L$     & $0.902$     \\
$M_{\rm NP}$ (TeV) & $~~~47.9~~~$     & $R(K)_C$     & $0.909$    \\
$~~~A_9~~~$          &  $~~~ -0.239~~~$    & $R(K^*)_L$   & $0.903$    \\
$~~~A_{10}~~~$     &  $~~~0.019~~~$    & $R(K^*)_C$   & $0.919$     \\
$~~~A_L\times 10~~~$   &  $~~~0.107~~~~$    & $~~~\Br(B_s\to\mu^+\mu^-)$  & $~~~~~3.59\times 10^{-9}$   \\
$~~~A_Q\times 10^2~~~$  &   $~~~ -0.991~~~$  & $~~~\Br(B^+\to K^+ \nu\nu)$  & $~~~~~2.31\times 10^{-5}$ \\
$~~~C_9^\NP~~~$   & $~~~ -0.370~~~$           &  $C_{10}^\NP$    &  $0.029$ 
\\ \hline\hline
\end{tabular}
\caption{Best-fit values of our fitting including $\Br(B^+\to K^+\mu^+\mu^-)$ and $\P5p$.}
\label{T_BF2}
\end{table}
%
The results are provided in Table \ref{T_BF2}.
The reduced $\chi^2_\min$ is $\chi^2_\min/\dof = 2.67$, 
and the fitting gets worse.
Compared to Table \ref{T_best_fit}, best-fit values of $R(\Ks)_{L,C}$ are slightly reduced
and that of $C_9^\NP$ gets negatively larger.
$1\sigma~(2\sigma)$ ranges of $C_{9,10}^\NP$ are
\begin{equation}
C_9^\NP = [-1.62, -0.492]~([-1.97, 0.684])~,~~~
C_{10}^\NP = [-0.599, 0.810]~([-0.719, 0.992]).
 \end{equation}
Tables \ref{T_BF_Kp} and \ref{T_BF_P5p} summarize our best-fit values and $2\sigma$ ranges of 
$\Br(B^+\to K^+\mu^+\mu^-)$ and $\P5p$.
%
\begin{table}
\begin{tabular}{c|cccc}
 $\Br(B^+\to K^+\mu^+\mu^-)\times 10^7~~~$   &   $~~~$ SM $~~~$ 
 &  $~~~~~$ experiment $~~~~~$  & best-fit values &    $2\sigma$ ranges \\\hline
 $[0.1, 0.98]$   &   $~~~ 0.32\pm0.03$     &   $0.29\pm0.02$   &  $0.24$     &    $[0.21,  0.27]$   \\
 $[1.1, 2.]$       &   $~~~ 0.33\pm0.03$     &   $0.21\pm0.02$   &  $0.25$     &    $[0.22,  0.27]$   \\
 $[2., 3.]$         &   $~~~ 0.37\pm0.03$     &   $0.28\pm0.02$   &  $0.27$     &    $[0.24,  0.30]$   \\
 $[3., 4.]$         &   $~~~ 0.37\pm0.03$     &   $0.25\pm0.02$   &  $0.27$     &    $[0.24,  0.30]$   \\
 $[4., 5.]$         &   $~~~ 0.37\pm0.03$     &   $0.22\pm0.02$   &  $0.27$     &    $[0.24,  0.29]$   \\
 $[5., 6.]$         &   $~~~ 0.37\pm0.03$     &   $0.23\pm0.02$   &  $0.26$     &    $[0.23,  0.29]$   \\
 $[6., 7.]$         &   $~~~ 0.37\pm0.03$     &   $0.25\pm0.02$   &  $0.26$     &    $[0.23,  0.28]$   \\
 $[7., 8.]$         &   $~~~ 0.38\pm0.04$     &   $0.23\pm0.02$   &  $0.25$     &    $[0.22,  0.28]$   \\
 $[15., 22.]$     &   $~~~ 1.15\pm0.16$     &   $0.85\pm0.05$   &  $0.72$     &    $[0.64,  0.80]$   \\        
\hline\hline
\end{tabular}
\caption{Best-fit values and $2\sigma$ ranges of our fittings 
for $\Br(B^+\to K^+ \mu^+\mu^-)$.
SM predictions and experimental data are from \cite{Alguero2304}.
}
\label{T_BF_Kp}
\end{table}
%
\begin{table}
\begin{tabular}{c|ccccc}
 $P_5'(B^+\to K^{*+}\mu^+\mu^-) ~~~$   &    $~~~$ SM $~~~$ 
 & $~~~~~$ experiment $~~~~~$  & best-fit values &    $2\sigma$ ranges \\\hline
 $[0.1, 0.98]$   &   $~~~ 0.68\pm0.14$       &   $0.51\pm0.32$    &  $0.67$      &    $[0.47,  0.88]$   \\
 $[1.1, 2.5]$     &   $~~~ 0.17\pm0.12$       &   $0.88\pm0.72$    &  $0.07$      &    $[-0.22,  0.48]$   \\
 $[2.5, 4.]$       &   $~~~ -0.45\pm0.11$     &   $-0.87\pm1.68$   &  $-0.53$     &    $[-0.87,  -0.05]$   \\
 $[4., 6.]$         &   $~~~ -0.73\pm0.08$      &   $-0.25\pm0.41$   &  $-0.76$     &    $[-0.95,  -0.21]$   \\
 $[6., 8.]$         &   $~~~ -0.81\pm0.07$      &   $-0.15\pm0.41$   &  $-0.84$     &    $[-0.94,  -0.34]$   \\
 $[15., 19.]$     &   $~~~ -0.57\pm0.05$      &   $-0.24\pm0.17$   &  $-0.65$     &    $[-0.69,  -0.37]$   \\   
\hline\hline
\end{tabular}
\caption{Best-fit values and $2\sigma$ ranges of our fittings 
for $P_5'(B^+\to K^{*+} \mu^+\mu^-)$.
SM predictions and experimental data are from \cite{Alguero2304}.
}
\label{T_BF_P5p}
\end{table}
%
\par
\begin{figure}
\begin{tabular}{cc}
\hspace{-1cm}\includegraphics[scale=0.12]{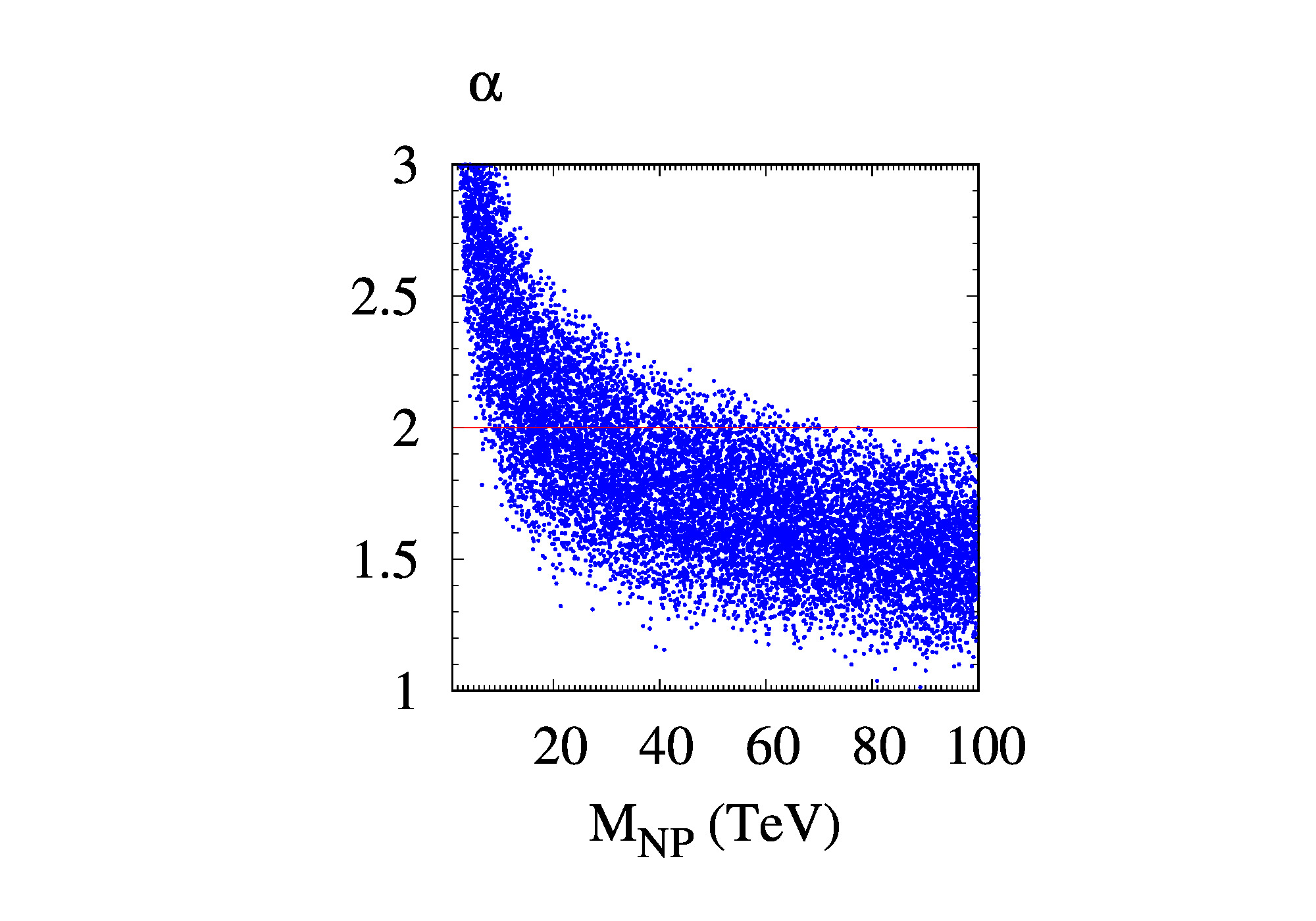} &
\hspace{-1cm}\includegraphics[scale=0.12]{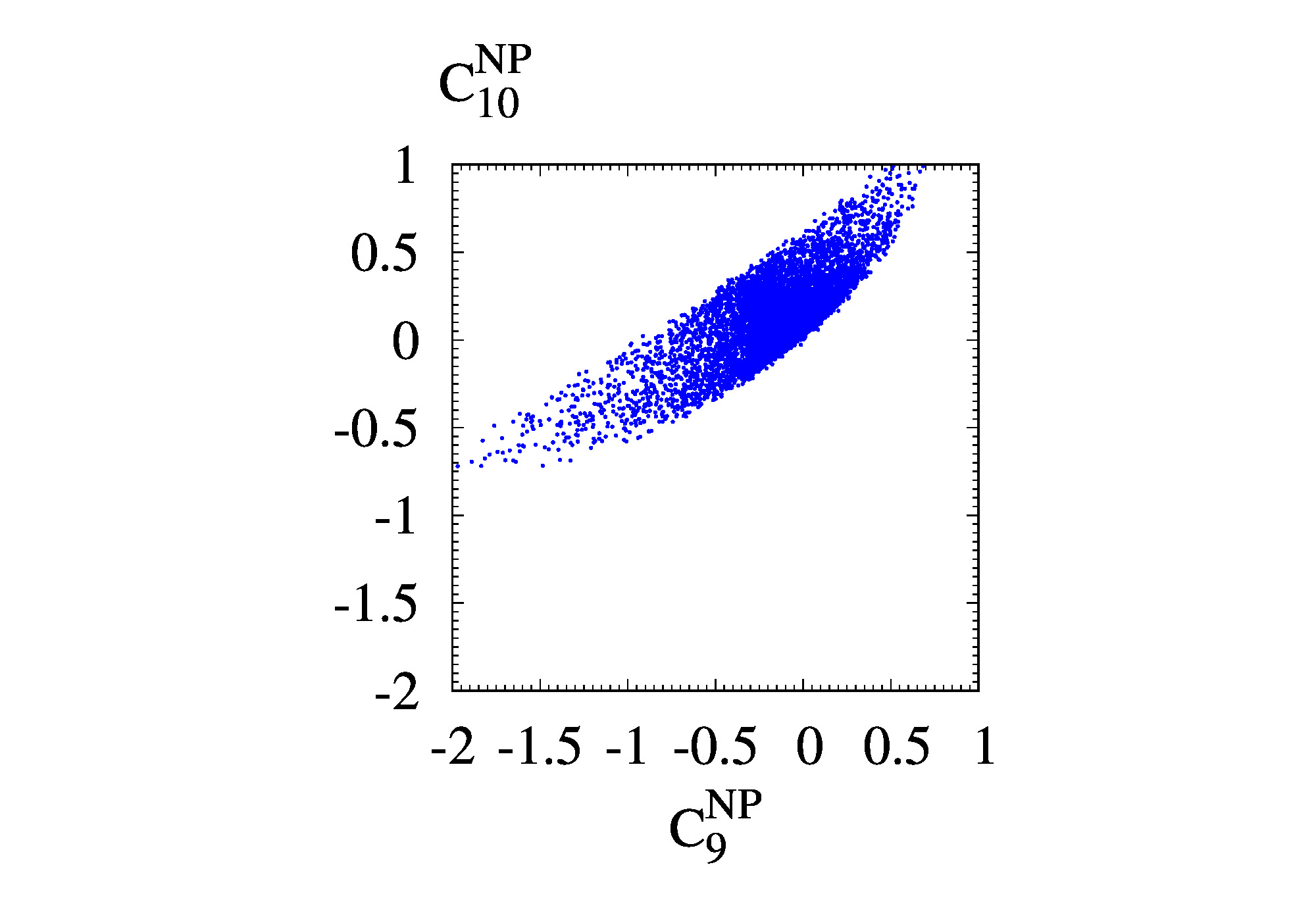} \\
(a) & (b) \\
\hspace{-1cm}\includegraphics[scale=0.12]{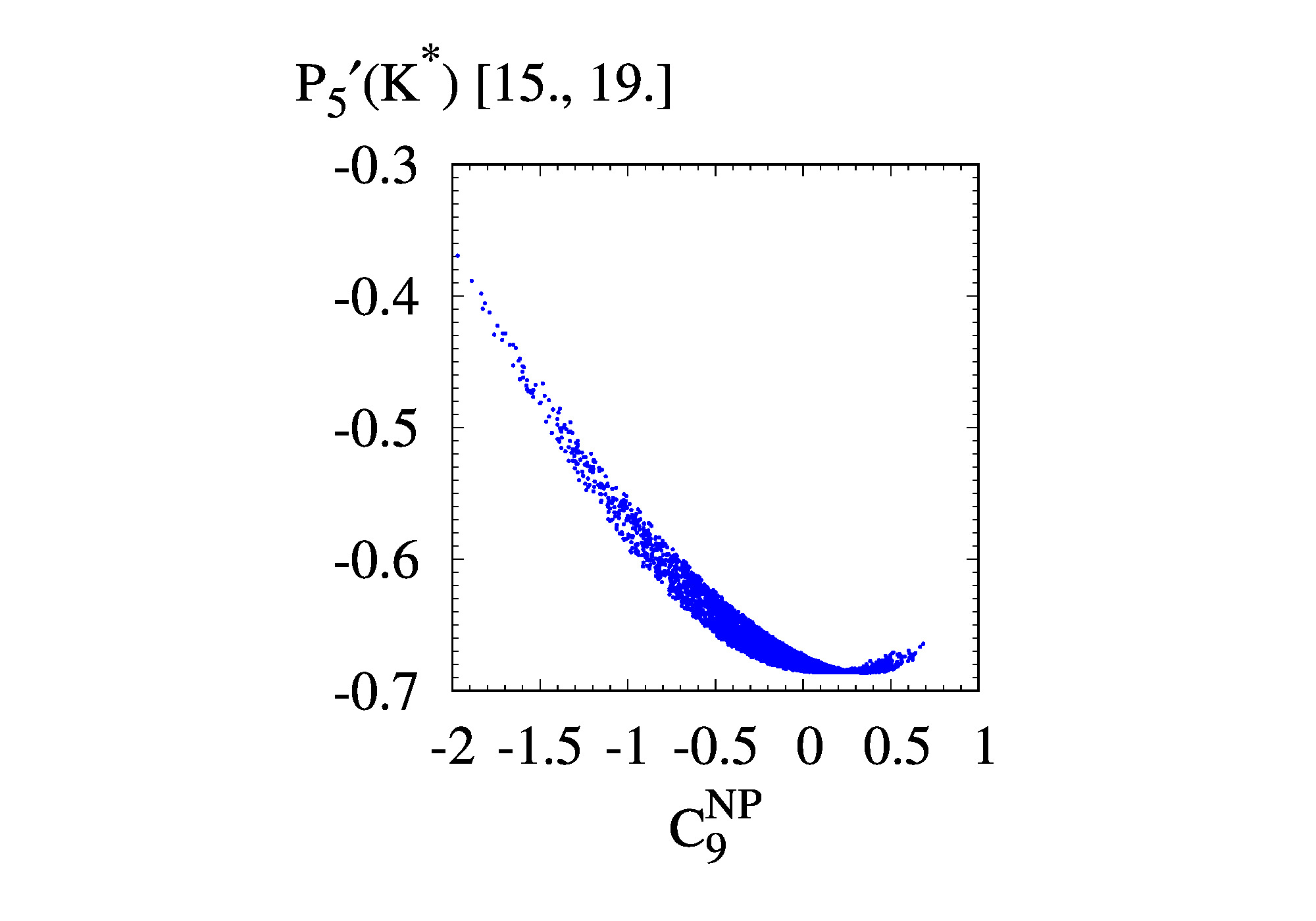} & 
\hspace{-1cm}\includegraphics[scale=0.12]{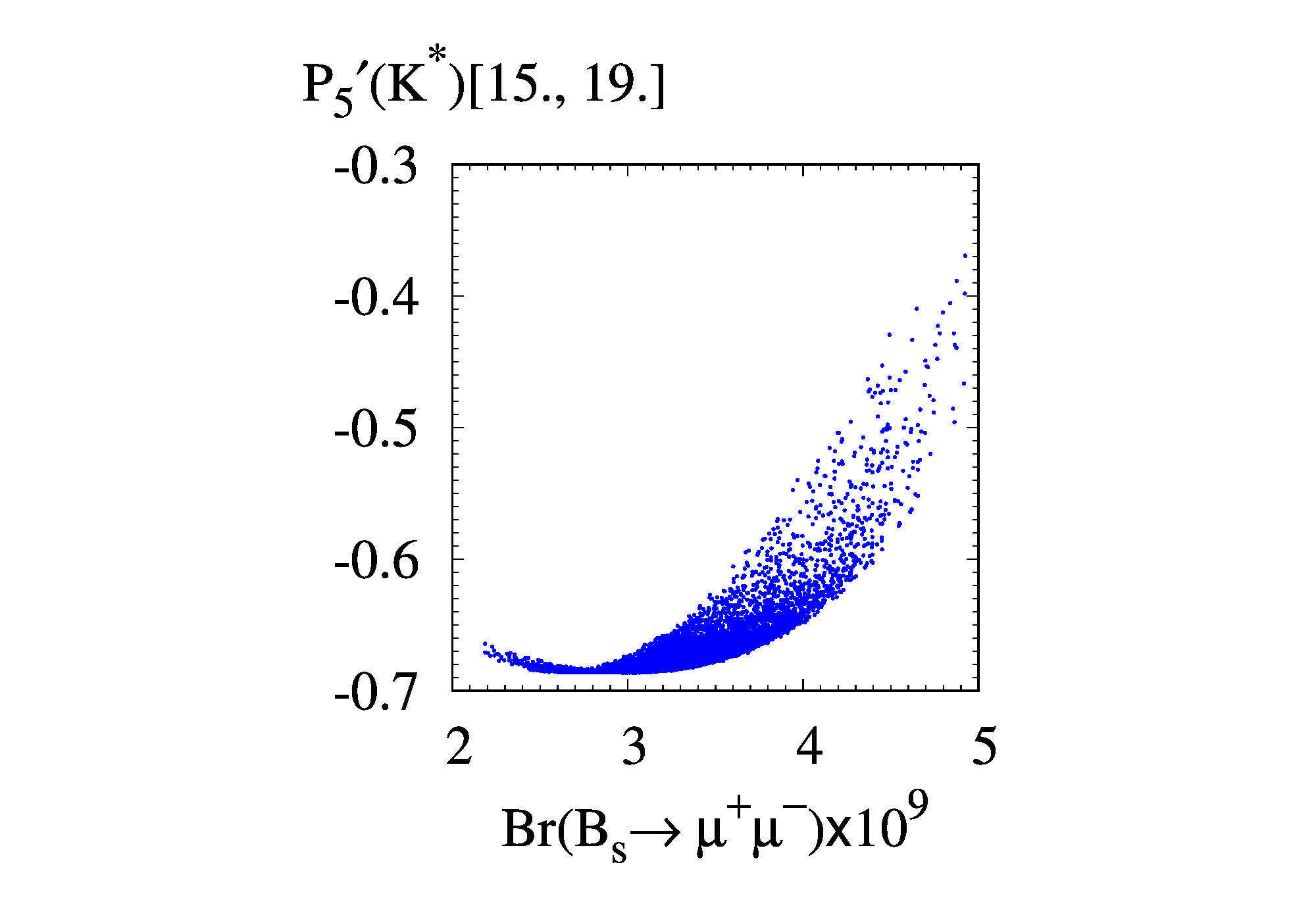} \\
(c) &  (d) \\
\end{tabular}
\caption{\label{fit_2} Allowed regions of 
(a) $\alpha$ vs. $M_\NP$,
(b) $C_{10}^\NP$ vs. $C_9^\NP$, 
(c) $P_5'(B^+\to K^{*+} \mu^+\mu^-)$ in $[15., 19.]$ bin ($P_5'(K^*)[15., 19]$ in short) vs. $C_9^\NP$, and
(d) $P_5'(K^*)[15., 19]$ vs. $\Br(\Bs2mumu)$
 at the $2\sigma$ level, with $\Br(B^+\to K^+\mu^+\mu^-)$ and $\P5p$ included.
}
\end{figure}

Figure \ref{fit_2} shows allowed regions of some parameters and observables at the $2\sigma$ level.
%
%
Compared to Fig.\ \ref{fit_para} (a) and (c), we have wider regions in the 
$\alpha-M_\NP$ and $C_{10}^\NP-C_9^\NP$ planes in Fig.\ \ref{fit_2} (a) and (b).
This is due to the larger $\chi^2_\min/\dof$.
The best-fit value of $M_\NP$ gets much higher, far beyond the reach of the HL-LHC.
For fixed $\alpha =2$ we have $4.66~\TeV\le M_\NP\le 83.7~\TeV$ at the $2\sigma$ level,
which is much wider window than the previous one.
Fortunately the exclusion mass limit of $Z'$ at the HL-LHC is still within the window.
Figure \ref{fit_2} (c) shows that larger values of $P_5'(K^*)[15., 19.]$ favor negatively large $C_9^\NP$.
For other bins $\P5p$ reveals similar patterns.
Also, Fig.\ \ref{fit_2} (d) illustrates that larger values of $P_5'(K^*)[15., 19.]$ require
larger $\Br(\Bs2mumu)$ than $\Br(\Bs2mumu)_\exp$, resulting in worse fitting.
It explicitly shows that $\Br(\Bs2mumu)$ exerts strong constraints on $\P5p$.
\par

%
%
%
%
%
%
\section{Conclusions}
%
%
%
In conclusion, we investigated possible NP effects on $b\to s$ transition processes.
Experimental data of the ratio $R(\Ks)$ and $\Br(\Bs2mumu)$ are compatible with the SM predictions
while $\Br(\B2Knunu)$ needs NP to explain the discrepancy of experiment and theory.
Considering these two kinds of data simultaneously turned out to be very interesting with regard to NP scale.
We parameterized NP effects in a generic way with NP scale $M_\NP$ explicitly 
with some possible power $\alpha$ of it.
Our best-fit value of $\alpha$ is less than 2 which corresponds to the case of ordinary new particles.
For a reasonable range of NP fermionic couplings we found that $3.17~\TeV\le M_\NP\le 14.9~\TeV$
for ordinary NP particles contributing at tree level.
The window of $M_\NP$ would vary for different ranges of couplings.
Since the width of the allowed window of $M_\NP$ is not so wide, 
our results would be very helpful to check validity of various NP scenarios in a generic way.
If new particles contribute via loop processes, then the NP scale would appear in a more complicated form,
which is beyond our current analysis.
\par
%
%
Experimentally, the HL-LHC is expected to exclude leptoquark masses up to $\lesssim 3~\TeV$,
slightly below our $M_\NP$ window.
But the $Z'$ mass limit at the HL-LHC is within the range.
\par
When including $\Br(B^+\to K^+\mu^+\mu^-)$ and $\P5p$, we found that the fitting gets worse,
$|C_9^\NP|$ can be $\calO(1)$,
and the upper bound of the $M_\NP$ window for $\alpha=2$ becomes much larger.
The lower bound of the window shifts a bit higher,  but still below the exclusion limit of the $Z'$ mass
at the HL-LHC.
\par
For specific NP models one should include constraints from $B^0\to K^{*0}\nu\nubar$ decays.
Experimentally we have an upper bound of $\Br(B^0\to K^{*0}\nu\nubar)<1.8\times 10^{-5}$ \cite{Belle1702}.
Future measurements will cast much more information for NP contributions.
%
%
%
\begin{acknowledgments}
This paper was supported by Konkuk University in 2024.
\end{acknowledgments}
%
%
%

%
\end{document}